\documentclass{emulateapj}
\usepackage{amsmath}
\usepackage{natbib}
\usepackage{multirow}
\usepackage{graphicx}
\usepackage{epstopdf}
\usepackage[backref,breaklinks,colorlinks,citecolor=blue]{hyperref}
\DeclareGraphicsExtensions{.eps,.png,.pdf}

\newcommand{\degree}{\ensuremath{^\circ}}

\newcommand{\Ha}{\ensuremath{\mathrm{H\alpha}}}

\shorttitle{The DECam Plane Survey}
\shortauthors{E. F. Schlafly et al.}
\begin{document}
\title{The DECam Plane Survey: Optical photometry of two billion objects in the southern Galactic plane}
\author{
E. F. Schlafly,\altaffilmark{1,2}
G. M. Green,\altaffilmark{3}
D. Lang,\altaffilmark{4,5}
T. Daylan,\altaffilmark{6}
D. P. Finkbeiner,\altaffilmark{6,7}
A. Lee,\altaffilmark{6}
A. M. Meisner,\altaffilmark{1}
D. Schlegel,\altaffilmark{1}
F. Valdes\altaffilmark{8}
}

\altaffiltext{1}{Lawrence Berkeley National Laboratory, One Cyclotron Road, Berkeley, CA 94720, USA}
\altaffiltext{2}{Hubble Fellow}
\altaffiltext{3}{Kavli Institute for Particle Astrophysics and Cosmology, Physics and Astrophysics Building, 452 Lomita Mall, Stanford, CA 94305, USA}
\altaffiltext{4}{Department of Astronomy \& Astrophysics and Dunlap Institute, University of Toronto, 50 Saint George Street, Toronto, ON, M5S 3H4, CA}
\altaffiltext{5}{Department of Physics \& Astronomy, University of Waterloo, 200 University Avenue West, Waterloo, ON, N2L 3G1, CA}
\altaffiltext{6}{Department of Physics, Harvard University, 17 Oxford Street, Cambridge, MA 02138, USA}
\altaffiltext{7}{Harvard-Smithsonian Center for Astrophysics, 60 Garden Street, Cambridge, MA 02138, USA}
\altaffiltext{8}{National Optical Astronomy Observatories, P.O. Box 26732, Tucson, AZ 85719, USA}

\begin{abstract}
The DECam Plane Survey is a five-band optical and near-infrared survey of the southern Galactic plane with the Dark Energy Camera at Cerro Tololo.  The survey is designed to reach past the main-sequence turn-off at the distance of the Galactic center through a reddening $E(B-V)$ of 1.5~mag.  Typical single-exposure depths are 23.7, 22.8, 22.3, 21.9, and 21.0 mag in the $grizY$ bands, with seeing around $1\arcsec$.  The footprint covers the Galactic plane with $|b| \lesssim 4\degree$, $5\degree > l > -120\degree$.  The survey pipeline simultaneously solves for the positions and fluxes of tens of thousands of sources in each image, delivering positions and fluxes of roughly two billion stars with better than 10 mmag precision.  Most of these objects are highly reddened and deep in the Galactic disk, probing the structure and properties of the Milky Way and its interstellar medium.  The full survey is publicly available.
\end{abstract}


\section{Introduction}
\label{sec:intro}

Many of the Milky Way's stars and much of its gas and dust reside in a disk.  Accordingly, observations of the Milky Way's disk are critical to understanding the Milky Way---particularly observations toward the inner Galaxy, where most of the mass lies.  At optical wavelengths, however, the interpretation of observations of the Milky Way's disk can be challenging due to the tremendous number of stars and due to extinction by dust, motivating surveys of the disk at infrared wavelengths where extinction is greatly reduced.

However, optical observations have advantages over infrared observations.  Critically, the intrinsic colors of typical stars depend more strongly on stellar type in the optical than in the infrared: optical colors are often more useful than infrared colors in determining the temperature of a star.  Additionally, for studies of the Milky Way's dust, the sensitivity of optical colors to extinction is valuable, allowing higher signal-to-noise measurements of the column of dust to stars.  Similarly, the optical extinction curve is more sensitive to variation in the properties of dust than is the infrared extinction curve, permitting the variation in the dust extinction curve to be studied throughout the Milky Way disk.

\begin{figure*}[htb]
\begin{center}\includegraphics[width=\textwidth]{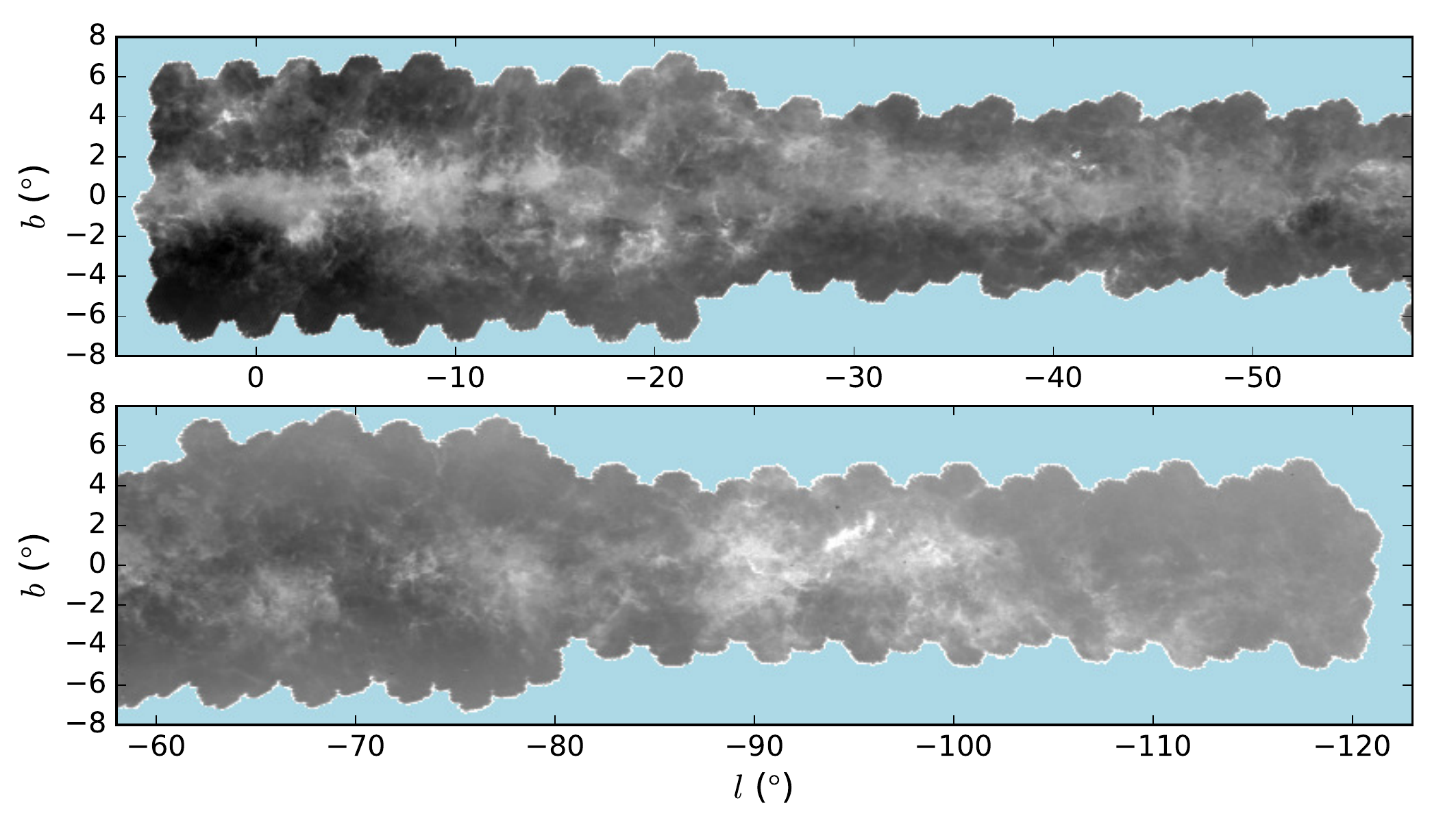}\end{center}
\caption[Source Density]{
\label{fig:sourcedensity}
The DECaPS footprint.  The figure shows the number density of sources detected in at least three bands, and brighter than 20th magnitude in $r$.  Dark regions have the most stars.  Major dust clouds are prominent, significantly reducing the number of stars observed behind them.  Prominent dust clouds include the Pipe Nebula at (1\degree, 4\degree) and the Vela Molecular Ridge cloud C at (-95\degree, 1.5\degree) \citep{Murphy:1991}.  
}
\end{figure*}

In this work, we present the DECam Plane Survey (DECaPS): $grizY$ imaging and derived catalogs of one third of the Milky Way's disk, covering roughly $|b| \lesssim 4\degree$, $5\degree > l > -120\degree$, as shown in Figure~\ref{fig:sourcedensity}.   Observations were made with the Dark Energy Camera (DECam) at Cerro Tololo, and are designed to reach the main sequence turn-off at a distance of 8 kpc through 1.5~mag $E(B-V)$.  The final survey catalog contains more than 20 billion photometric measurements of two billion objects, a rich trove of data for understanding the Milky Way and its stars and dust.

DECaPS occupies a special niche among the many surveys targeting the Milky Way.  The most comparable survey is the PS1 survey \citep{Chambers:2016}.  DECaPS and PS1 use a very similar set of filters and achieve similar depths (DECaPS is roughly 1 magnitude deeper in individual images), though PS1 covers the entire sky above $\delta = -30\degree$, and contains more epochs than DECaPS does (12 vs. 3).  PS1 has been a valuable survey for understanding the structure of the Milky Way's disk, both in stars \citep{Slater:2014, Morganson:2016} and dust \citep{Green:2015, Schlafly:2015, Hanson:2016, Schlafly:2016, Schlafly:2017}.  DECaPS intends to duplicate PS1's success in the southern Galactic plane sky where no PS1 imaging is available, while obtaining greater depths to reach further across the Milky Way's disk.

Other optical surveys targeting the Milky Way's disk include IPHAS \citep{Drew:2005} and VPHAS+ \citep{Drew:2014}, which are roughly a magnitude shallower than DECaPS.  In contrast to DECaPS, however, these surveys include observations in \Ha\ (and $u$ in VPHAS+), critical for identifying ionized nebulae and unresolved pre- and post-main sequence stars.

Near-infrared surveys like 2MASS \citep{Skrutskie:2006} have long been used to study the Galactic plane, in part because of the near-infrared's reduced sensitivity to extinction relative to the optical.  Dedicated surveys of the Galactic plane reach dramatically fainter magnitudes, like the UKIDSS Galactic Plane \citep{Lucas:2008} and Vista Variables in the Via Lactea surveys \citep{Minniti:2010}.  Finally, space missions like the all-sky WISE survey \citep{Wright:2010} and the plane-focused Spitzer GLIMPSE surveys \citep{Benjamin:2003, Churchwell:2009} have imaged the entire Galactic plane at longer wavelengths, where dust extinction has a still smaller effect.

The combination of surveys at different wavelengths offers the best approach to disentangling the rich phenomena at play in the inner Galaxy, providing long-wavelength observations that are relatively insensitive to dust and pierce deep into the Milky Way's enshrouded disk, as well as optical observations that are much more sensitive to both dust and to intrinsic stellar properties, but can be effectively much shallower in regions of high dust column.  DECaPS provides optical observations of the southern Galactic plane to complement observations at longer wavelengths to enable to new maps of stars and dust in the southern Galactic plane.

This paper is organized into 8 sections.  In \textsection\ref{sec:intro} and \textsection\ref{sec:strategy}, we introduce DECaPS and its strategy.  In \textsection\ref{sec:pixel} and \textsection\ref{sec:catalog}, we describe the pixel-level processing of the images and their reduction into catalogs.  Next we describe the photometric calibration in \textsection\ref{sec:cal}.  The characteristics of the survey are described in \textsection\ref{sec:characteristics}.  Information about the data release is provided in \textsection\ref{sec:release}, and we conclude in \textsection\ref{sec:conclusion}.

\section{Survey Strategy}
\label{sec:strategy}

The DECam Plane Survey is a survey of the southern Galactic plane in five broadband filters from 400~nm to 1050~nm ($grizY$) in arcsecond seeing, to depths of 23.7, 22.8, 22.3, 21.9, and 21.0 mag, respectively.  The survey covers $\delta < -30\degree$, $|b| \lesssim 4\degree$, about 1000 square degrees.  More than two billion sources are detected; the spatial distribution of bright sources is shown in Figure~\ref{fig:sourcedensity}.  When combined with PS1, the two surveys cover the entire Galactic plane with arcsecond-resolution five-band optical photometry.  The basic parameters of the survey are given in Table~\ref{tab:basicparam}.

\begin{deluxetable}{ccccccc}
\tablewidth{\columnwidth}
\tablecaption{DECam Plane Survey Parameters}
\tablehead{
\colhead{Filter} & \colhead{$\lambda$} & \colhead{Time} & \colhead{6$\sigma$ Depth} & \colhead{Num} & \colhead{FWHM} & \colhead{Sky}
\label{tab:basicparam}
}
\startdata
$g$ & 480 & 96 & 23.7 & 3 & 1.16 & 21.4 \\
$r$ & 638 & 30 & 22.8 & 3 & 1.07 & 20.6 \\
$i$ & 777 & 30 & 22.3 & 3 & 0.97 & 19.3 \\
$z$ & 911 & 30 & 21.9 & 3 & 0.93 & 18.7 \\
$Y$ & 985 & 30 & 21.0 & 3 & 0.91 & 18.0 \\
\enddata
\tablecomments{
Basic parameters of DECaPS.  The survey delivers photometry from about 400 to 1050~nm in 5 broad bands, to depths between 21st and 24th magnitude with seeings around an arcsecond.  Each part of the footprint was observed roughly three times in photometric conditions.  The table gives the filters observed in, their central wavelengths in nm, the typical exposure times, the $6\sigma$ limiting magnitudes, the typical number of photometric observations of each part of the footprint, the full-width at half-maximum (FWHM) in arcseconds (0.262\arcsec\ pixels), and the sky brightness in mag per square arcsecond.
}
\end{deluxetable}

DECam is the best instrument available for this survey.  The survey requires both a large aperture and a large field of view to efficiently survey the faint, extinguished stars in the Galactic plane.  The Blanco and DECam deliver the excellent seeing and image quality required to resolve the billions of stars in the often extremely crowded inner Galaxy.

\subsection{Depth}
\label{subsec:depth}
DECaPS is designed to measure the fluxes stars over a substantial fraction of the Milky Way's disk.  Specifically, the survey targets a depth reaching below the main-sequence turn-off at a distance of eight kiloparsecs.  For unextinguished stars, this would correspond to a magnitude of roughly 18.5, but substantial extinction in the Galactic plane requires dramatically deeper observations.  Extinction in the Galactic midplane varies greatly; much of the outer Galaxy has $E(B-V) < 1$~mag, but, especially in the inner Galaxy, extinction can become very large.  Through $E(B-V) = 1.5$~mag at 8~kpc, the main sequence turn-off lies at 24.1, 22.3, 21.3, 20.6, and 20.4 mag.  DECam can reach these magnitudes at $5\sigma$ in less than 30 second exposures in the $rizY$ bands, while in $g$ band 96 seconds are required to reach our target depth.  Because overheads between exposures with DECam are close to 30 seconds, it is wasteful to take exposures much shorter than 30 seconds; accordingly, we adopt a 96 second exposure time in $g$ and 30 seconds in all other bands.  Given the conditions actually obtained at the telescope, the survey reached typical 6$\sigma$ depths of 23.7, 22.8, 22.3, 21.9, and 21.0 mag in $grizY$.  The extra depth in the redder bands improves our distance reach in extinguished regions where $g$ photometry, in particular, is not practical.  The survey is slightly shallow in individual exposures in $g$, but three images are available of each part of the sky, allowing improved average photometry.

\begin{figure}[htb]
\begin{center}\includegraphics[width=\columnwidth]{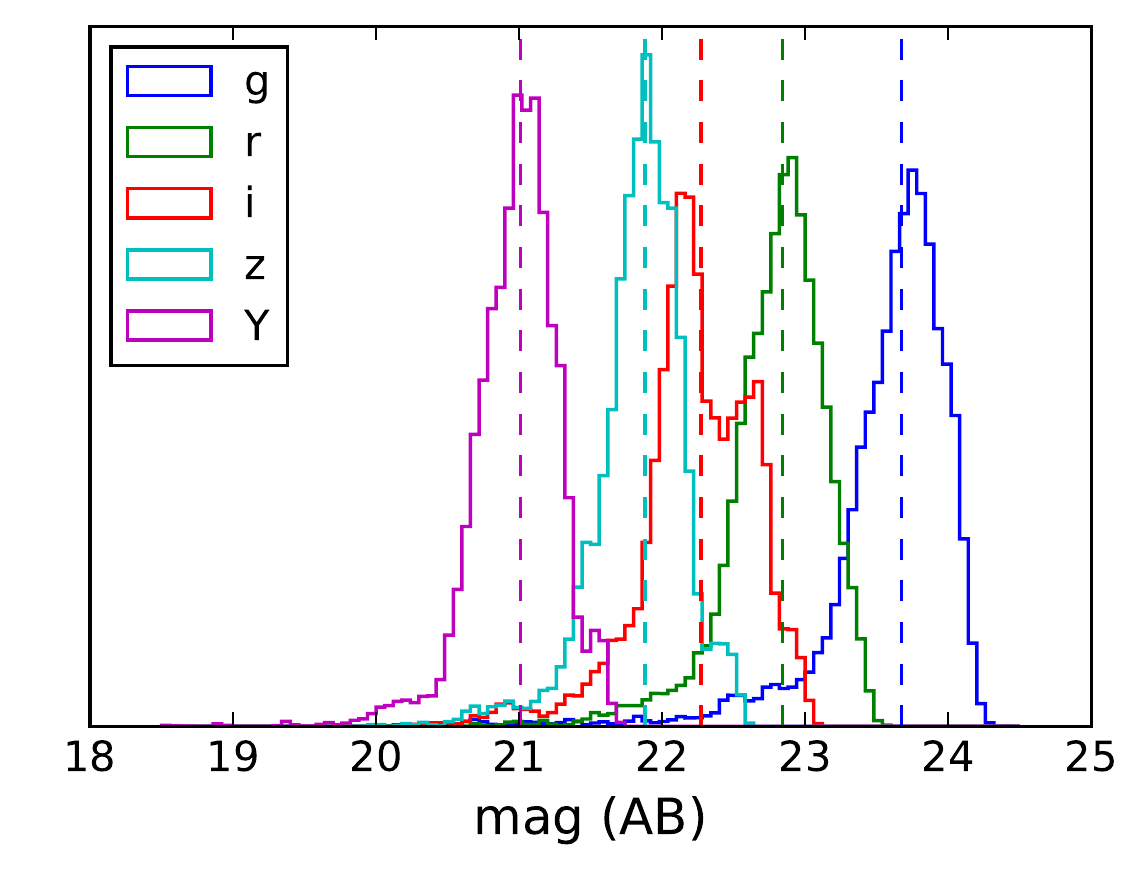}\end{center}
\caption[Depth Histogram]{
\label{fig:depths}
The distribution of individual image depths in the DECaPS survey.  Distributions are tight, giving homogeneous data over the survey footprint.  Means of the distributions are indicated by vertical dashed lines.
}
\end{figure}

Figure~\ref{fig:depths} shows the 6$\sigma$ depths obtained by DECaPS over all exposures taken by the program.  The distributions of depths in each filter are tight.  A few shallower exposures in each filter are attributable to cloudy weather.  The $i$ band depth distribution is bimodal due to significantly darker skies in the second half of the DECaPS observations than in the first.

\subsection{Tiling}
\label{subsec:tiling}
The survey aims for coverage of the entire $|b| \lesssim 4\degree$, $\delta < -30\degree$ sky.  A small fraction of the DECam focal plane is not available for imaging, primarily due to gaps between the CCDs.  To address this problem, DECaPS adopts a three-pass dither strategy, where each part of the footprint is observed with three different exposures, slightly offset from one another.  The survey uses the three-pass tiling strategy developed for the DECam Legacy Survey \citep{DESI:2016a, DESI:2016b}.  

This tiling scheme uniformly covers the sky with pointings separated by distances closely matched to the DECam field of view.  The field centers of the three passes are slightly rotated relative to one another to dither the exposures around each tile center, filling in the gaps in the DECam focal plane.

The DECaLS tiling scheme happens to contain dithers primarily in the right-ascension direction around right ascensions of $90\degree$ and $270\degree$.  This leads to a small amount of area without any DECaPS coverage within 10\degree\ of the Galactic center.

\subsection{Cadence}
\label{subsec:cadence}
The DECaPS cadence aims to observe each part of the footprint in photometric conditions on three separate nights.  Typically these three nights fall in one observing run and occur subsequently, but the vagaries of the weather led to a few regions in which two passes were observed immediately after one another on a single night, and others in which repeat observations of the same field were taken about one year apart from one another.

This cadence was designed to enable an internal photometric calibration based on repeated observations of the same stars \citep[e.g.][]{Padmanabhan:2008, Schlafly:2012, Burke:2017}.  Observing in different passes on different nights then both filled in the gaps in the footprint that would be left by a single pass, and provided the repeat observations needed to constrain the atmospheric throughput as a function of time.

\subsection{Observations}
\label{subsec:observations}
We obtained 22 nights on the Blanco 4m through the NOAO.  The observing runs are described in Table~\ref{tab:observations}.  Observations were made in $gr$ when the moon was down and $izY$ when the moon was up.  The longer exposure time in $g$ meant that the survey needed roughly equal amounts of moon-up and moon-down time.  The efficiency of the DECam system was excellent, enabling observations to keep up with the footprint as it crossed the meridian.  The southern location of the footprint and telescope, combined with favorable scheduling, allowed the survey to obtain very low airmasses, with a mean airmass of 1.15 and a standard deviation of 0.10.

\begin{deluxetable}{crll}
\tablewidth{\columnwidth}
\tablecaption{Observing Runs}
\tablehead{
\colhead{Start} & \colhead{\# Nights} & \colhead{Filters} & \colhead{Notes}
\label{tab:observations}
}
\startdata
2016-03-13 & 4 & $gr$ & 2016-03-17 clouded out \\
2016-03-23 & 4 & $izY$ & 2016-03-25 clouded out \\
2016-08-10 & 0.5 & $izY$ & Scattered clouds \\
2016-08-14 & 0.5 & $izY$ & Scattered clouds \\
2016-08-15 & 0.5 & $izY$ &  \\
2016-08-22 & 5$\times$0.5 & $gr$ & 2016-08-23 \& 25 clouded out \\
2017-01-16 & 8$\times$0.5 & $grizY$ & 2017-01-22 \& 23 clouded out \\
2017-04-19 & 2 & $grizY$ & 2017-04-19 marginal \\
2017-04-27 & 2 & $gr$ &  \\
2017-04-29 & 2$\times$0.5 & $gr$ &  \\
2017-05-03 & 2$\times$0.5 & $grizY$ & 2016-05-04 cloudy
\enddata
\tablecomments{DECaPS observing runs.  The runs included 12 full nights and 20 half nights, most of which were clear.  The number of nights in each run and the primary filters observed on those nights are listed, together with notes about any inclement conditions.
}
\end{deluxetable}

Due to a slight imbalance in the amount of dark and bright time available in good conditions, $r$ band observations in January 2017 were taken with longer 50 sec exposures when the moon was up, and in April/May 2017 some $izY$ exposures were taken with the moon down.

\section{Pixel Processing}
\label{sec:pixel}

DECaPS constructs catalogs from calibrated images delivered by the NOAO Community Pipeline \citep[CP]{Valdes:2014}.  The CP transforms raw images into calibrated, science-grade data, performing the following steps (among others):
\begin{enumerate}
\item bias subtraction,
\item crosstalk correction,
\item flat fielding,
\item sky subtraction,
\item artifact masking (cosmic, satellite, saturation, bleed trail),
\item weight image construction,
\item astrometric solution, and
\item initial photometric calibration.
\end{enumerate}
The DECaPS pipeline uses the CP calibrated images (\texttt{InstCal}), as well as the associated weight maps and data quality images.  Any pixel marked in the CP data quality image as being problematic, except for pixels marked as being ``transient,'' are given zero weight in subsequent analysis.

The CP data products were well suited for analysis with the DECaPS pipeline.  The DECaPS pipeline performed only a few additional pixel-level corrections and additions:
\begin{enumerate}
\item supplementary cosmetic masking,
\item removal of jump in background level on CCD S7,
\item and identification of nebulosity.
\end{enumerate}
We describe these additional steps in some detail as they are unique to DECaPS, though they represent only a small amount of processing relative to the CP.

\subsection{Cosmetic Masking}
\label{subsec:masking}
While analyzing CP images we found a number of bad columns caused by charge traps in the CCDs.  In the worst cases, these were very significant and could lead the DECaPS pipeline to detect hundreds of spurious sources along the bad columns.  To identify these features, we high-pass filtered images by subtracting a 7$\times$7 median filter.  We then took the median value of each pixel in a stack of 100 high-pass filtered images to identify pixels that were consistently significantly high or low.  This process identified 24,357 pixels that had not been identified as problematic by the CP and were consistently discrepant by more than 8 counts.  This is a vanishingly small fraction of the DECam focal plane ($\approx 0.005\%$), but in low-density regions these pixels could result in large numbers of spurious sources.  This improved masking has since been incorporated directly into the CP.

\subsection{S7 background level}
\label{subsec:s7}

The gain on amplifier B on DECam CCD S7 is unstable.  However, the CCD is often well behaved, motivating preservation of this CCD if possible.  Significantly discrepant gain, however, can lead to a large jump in sky level at the amplifier boundary, which the DECaPS pipeline sky algorithm does not track.  The result can be large positive backgrounds that the pipeline will try to explain as a blend of a large number of stars.

To avoid this, on S7 only, the pipeline compares the 10 pixels on each side of the amp boundary with one another.  A linear function of row number is robustly fit to the difference between pixels and their mirrored pixels across the amp boundary.  The derived fit is then applied to all pixels in amplifier B on S7 to remove the jump in background level.  This by no means fully accounts for potential problems on S7 (a multiplicative problem has been ``corrected'' with an additive offset), but it is adequate to prevent catastrophic failure and to enable downstream processing to identify and exclude S7 amplifier B.  For precision photometry in the merged DECaPS catalogs, S7 amplifier B is always excluded.

\subsection{Nebulosity}
\label{subsec:nebulosity}

The DECaPS pipeline attempts to explain all flux in images as coming from a smooth background (i.e., the sky) and stars. This fails badly in the presence of galaxies and nebulosity, which the pipeline attempts to shred into many stars. For the most part, it seems an adequate simplification to shred galaxies into stars in the DECaPS footprint: the vast majority of the sources \textit{are} stars, and both the algorithmic challenge and computational complexity of modeling galaxies is not justified. Nebulosity is more problematic. A small fraction ($\sim$0.1\%) of the footprint features substantial diffuse emission, primarily in the form of H$\alpha$, \ion{O}{3}, and scattered light from dust. This nebulosity can contain substantial fine structure that is not compatible with the DECaPS smooth sky model, leading the pipeline to try to explain the nebulosity as the sum of thousands of carefully distributed stars.

Convolutional neural networks are ideally suited to the task of recognizing nebulosity in an image. Convolutional neural networks (CNNs) have found wide use in image, signal and natural language processing, among other fields (see \citealt{LeCun2015} for an accessible introduction). CNNs process input in a way that respects locality (e.g., nearby pixels in an image are more closely related than more distant pixels) and translation invariance (e.g., a given object might appear anywhere in an image). CNNs work by first identifying features on small scales (by sliding different convolutional filters across the input), and then connecting these features on ever larger scales to build up a rich representation of the input. In a typical classification problem, where one has labeled training data, one first defines the overall structure of the CNN (the number of layers in the network, the number of convolutional filters to include in each layer, etc.). Then, one feeds training data into the network, compares the network output with the desired answer, and uses backpropagation to vary the weights in the network so that the residuals are reduced. After many training epochs, the network ``learns'' the convolutional filter weights that minimize the classification errors. CNNs have proven remarkably successful in solving image- and signal-processing problems that humans can intuitively solve by eye or by ear (e.g., facial recognition, speech recognition, or identifying cats in images). A human with little to no prior knowledge can be quickly trained to identify nebulosity in astronomical images, making CNNs an obvious approach to solving this problem.

We trained a CNN to identify $512 \times 512$-pixel regions of images containing significant nebulosity.  After identification, these regions were flagged by adding a bit to the CP data quality images (Table~\ref{tab:flags}).  Downstream in the analysis pipeline, putative stars in these regions were required to satisfy a sharpness criterion and be relatively uncontaminated by light from any nearby stars; otherwise the putative stars were eliminated from the model. This prevents the pipeline from trying to explain the flux in large nebulous regions with thousands of point sources, at the expense of reducing the pipeline's capability to model blended stars in these regions.

In order to provide training and validation data for our network, we hand-classified $512 \times 512$-pixel images. We sorted the images into four categories:
\begin{enumerate}
    \item \texttt{NEBULOSITY} -- significant contamination by nebulosity.
    \item \texttt{NEBULOSITY\_LIGHT} -- faint nebulosity.
    \item \texttt{NORMAL} -- no contamination.
    \item \texttt{SKY\_ERROR} -- spurious fluctuations in sky level injected by upstream DECam pipeline.
\end{enumerate}
We arrived at a dataset of 2000 images labeled \texttt{NORMAL}, 1775 images labeled \texttt{NEBULOSITY}, 1058 images labeled \texttt{NEBULOSITY\_LIGHT} and 629 images labeled \texttt{SKY\_ERROR}. We used 80\% of the images to train the network, and the remaining 20\% to validate the network. This is a relatively small number of training images with which to train a CNN. We therefore augmented our dataset by flipping the images vertically and horizontally, but did not perform other augmentation techniques (such as arbitrary rotation) that might alter the noise properties of the images or remove valuable information (e.g., the orientation of diffraction spikes). We histogram-normalized the images before feeding them to the neural network, which ensures that the sky background is always discernible.

On the validation dataset, our trained network achieved 90\% completeness and 90\% purity in its \texttt{NEBULOSITY} classifications, with a vanishingly small percentage of \texttt{NORMAL} images being mis-classified as \texttt{NEBULOSITY}. Our final validation loss (a measure of the accuracy of the classifications) was similar to our training loss, indicating that our network does not suffer from over-fitting.

We applied our trained convolutional neural network to each $512 \times 512$-pixel region of each survey image in order to flag areas with nebulosity. The neural network identified regions affected by significant nebulosity very accurately. A few corner cases were marked as nebulous, such as artifacts near extremely bright stars (brighter than 6th mag), or ghosts associated with these stars. These regions are extremely rare, and from the perspective of the catalog, it is appropriate to flag them as nebulous anyway: the smooth sky plus stars modeling is likewise inadequate here.

A more technical description of our CNN is provided in Appendix~\ref{app:nebulosity-network-structure}.

\subsection{Failure modes}
\label{subsec:pixfailure}

The resulting images after CP processing are an extremely clean description of the sky, with most instrumental signatures accurately removed.  A few remaining problems impact a small fraction of the images, however.

First, in regions featuring very significant nebulosity (primarily $\eta$ Carinae), the CP sky subtraction has difficulties and introduces artifacts into the images.  Fortunately, these artifacts are spatially smooth and are absorbed into the DECaPS pipeline sky estimates, and so these artifacts are not expected to influence the source detection or photometry in DECaPS.  They do, however, lead to unsightly regions in DECaPS images in these regions.

Second, the community pipeline has difficulties identifying cosmic rays in crowded regions.  Moreover, cosmic rays that are identified are frequently only partially masked.  This leads to occasional spurious sources in the DECaPS source catalogs.  Fortunately, the relatively short DECaPS exposures and dense fields means that only a tiny number of sources are affected.  Nevertheless, sources detected in only a single image, especially when partially masked (\texttt{QF $<$ 1}, Table~\ref{tab:individualcatalog}), are likely to be spurious.

Bleed trails associated with very bright stars occasionally introduce crosstalk artifacts in images.  In rare cases on DECam CCD N15, the crosstalk correction for extremely saturated parts of the bleed trail is incorrect and leads to dramatic oversubtraction of the crosstalk on the adjacent amplifier.  This case introduces spurious sources in the DECaPS processing, because the only mechanism the DECaPS modeling has to address regions with substantial negative flux is to reduce the sky level.  This can leave significant positive sky on the image everywhere away from the crosstalk artifact, corrupting the flux measurements of stars throughout the image and potentially introducing numerous spurious stars that attempt to explain the high background.  Fortunately, we have only observed this effect on a few individual CCD frames of the 600,000 composing the survey.

Related effects occur whenever processing errors leave large negative fluxes in the images.  Very bright stars saturate the serial register, leading to large artifacts in the images.  These artifacts are usually masked by the CP, but in rare cases they are not masked and result in spurious sources and contaminated fluxes.

\section{Catalog Construction}
\label{sec:catalog}
DECaPS catalogs were constructed by the \texttt{crowdsource} crowded field photometry pipeline, which is described in detail in a future paper.  These catalogs were then organized and merged using the Large Survey Database package \citep{Juric:2012}.  

Briefly, the \texttt{crowdsource} pipeline models images as the sum of a smooth sky component plus the sum of a large number of sources that share a common point-spread-function (PSF) description, similar to the approaches taken by DAOPHOT \citep{Stetson:1987}, PS1 \citep{Magnierdet:2016}, DOLPHOT \citep{Dolphin:2000}, and the Tractor \citep{Lang:2016}.  The pipeline takes advantage of the \texttt{LSQR} sparse least squares routine \citep{Paige:1982} to simultaneously solve for the sky level as well as the fluxes and positions of thousands of sources.  The densest fields in the survey feature 30,000 sources per $1024\times1024$ pixels, leading to 90,000 parameters in the linear least squares fits (flux, $x$, and $y$ coordinates for each of 30,000 sources).

\texttt{Crowdsource} iteratively finds fainter, more blended sources, and improves its PSF and sky estimates.  Each iteration consists of the following steps:
\begin{enumerate}
  \item Smooth sky subtraction
  \item Source detection on residual image
  \item Fit fluxes, positions, and global sky level
  \item Improve positions via centroid computation
  \item Improve PSF
\end{enumerate}

Each iteration begins with the current best model image and PSF.  In the initial iteration, the best model image is simply constant and equal to zero, and the PSF shape is determined from the CP-measured full width at half maximum (FWHM).

After each iteration, a new PSF model, source list, and improved source locations are available, which are used as input to the next iteration.  The pipeline continues iterating until at least four iterations have been performed.  Afterward, the pipeline will cease iterating if fewer than 100 new sources have been detected, or if ten iterations have occurred.  The ten-iteration limit is reached in very dense regions like Baade's window, and very rarely elsewhere.  Finally, a last pass is made without detecting new sources or allowing positions to vary; only forced photometry at the existing list of stellar positions is performed.

\begin{figure*}[htb]
\begin{center}\includegraphics[width=\textwidth]{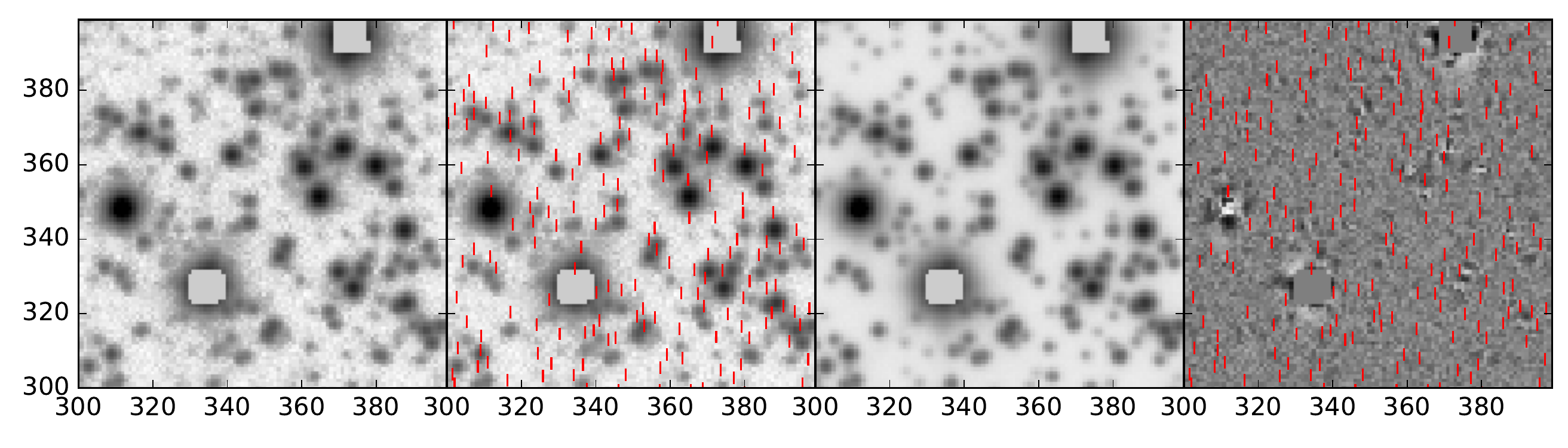}\end{center}
\caption[Image-Model Comparison]{
\label{fig:immodcomp}
Comparison between a 100$\times$100 pixel portion of a $z$-band image in a dense region and the \texttt{crowdsource} model for that image.  The first panel shows the image; the second panel the image with detected sources overlaid; the third panel the model image; and the fourth panel the residual image.  The red lines are centered five pixels above each detected source.  The pipeline does an excellent job detecting and photometering the sources in the image, resulting in a model that well describes the observation.  The image shown represents about $2 \times 10^{-9}$ of the total DECaPS imaging.
}
\end{figure*}

In general, this procedure is very successful.  Figure~\ref{fig:immodcomp} examines a 100$\times$100 pixel region from a $z$-band image in the vicinity of $(l, b) = (-9.7\degree, 3.3\degree)$: towards the inner Galaxy and somewhat off the midplane where the most extinction lies.  The first panel shows the image with an arcsinh stretch; the second panel shows the detected sources; the third panel shows the model image; and the fourth panel shows the residual image after additional stretching.  Saturated regions flagged by the CP are set to zero.  Imperfections in the PSF model lead to significant coherent residuals around all bright stars.  However, the model does an excellent job identifying and attributing fluxes to blended sources; many complicated asterisms are neatly resolved into individual stars.  For example, the stars near (395, 315) are neatly identified and photometered, though one faint star goes unidentified.  The pipeline tries to be conservative when deblending sources.  A number of close blends are not deblended and are readily recognized in the residual image as pairs of dark, positive residuals flanking a central light, negative residual.  For instance, one such case is evident near (345, 375).

Ultimately \texttt{crowdsource} relies on simple, rigid heuristics to determine identify a fixed set of objects in a field.  More principled, probabilistic techniques are possible, and can better accommodate the highly covariant fluxes of different sources, albeit at greater computational expense.  Some such techniques are explored in \citet{Brewer:2013}, \citet{Daylan:2017} and \citet{Portillo:2017}.

\subsection{Smooth Sky Subtraction}
\label{subsec:smoothsky}
In the initial smooth sky subtraction, the median of each 20$\times$20 pixel subregion of the residual image (image minus best-fit model image) is computed.  The pipeline linearly interpolates between these median values to construct a smooth sky image, which is subtracted from the residual image.  This sky estimate will be biased high by any undetected sources in the image.

\subsection{Source Detection}
\label{subsec:sourcedetection}
Next, sources are detected.  The pipeline smooths the residual image with the PSF and identifies peaks with greater than $5\sigma$ significance, according to the CP inverse variance image.  This is equivalent to a matched-filter analysis.  Peaks that are too blended with stars in the current best fit model image are excluded.  Peaks are kept if the significance of the peak is more than 40\% of the significance of the model image at that location, or if both the peak flux and the peak significance is more than 20\% of the model flux and significance.

\subsection{Flux and Position Solution}
\label{subsec:fluxpossol}
Any stars found are added to the list of currently detected stars in the image.  These new stars may influence the fluxes and positions of other stars, at least through their influence on the sky, as well as through blending.  \texttt{Crowdsource} simultaneously solves for an overall sky level and the fluxes and positions of all of the currently identified stars using the linear least squares routine \texttt{LSQR}.  A source's image is not a linear function of its position.  Accordingly, for use with the linear least squares routine we linearize the problem by approximating changes in the position of the source by the first term in the Taylor series expansion.  Fitting an overall sky level allows the fit to partially account for the initially high sky estimate.

\subsection{Centroiding}
\label{subsec:sky}
The positions estimated in \textsection\ref{subsec:fluxpossol} are linearized approximations to the positions, and are accurate only for small changes in the positions of the source.  Better performance can be obtained by computing the sources' centroids, but this is challenging in a crowded field, where neighboring stars' flux is blended with a target star's flux.  \texttt{Crowdsource} attempts to overcome this by using its new best-fit model of the image to eliminate contaminating neighboring stars from the centroid computation.  The pipeline visits each source in turn, and subtracts its model of all other sources from the image.  The source's centroid is then measured on the neighbor-subtracted image.  These centroids are used as input positions in the next iteration.

\subsection{PSF Modeling}
\label{subsec:psf}
Finally, the PSF model is improved.  The PSF is modeled as an ideal-seeing PSF, convolved with an elliptical Moffat PSF \citep{Moffat:1969} representing the current seeing, plus a $9\times9$ residual image in the core of the PSF.  The parameters of the Moffat PSF, and the value of each pixel of the $9\times9$ residual image are allowed to change linearly over each $1024\times1024$ subregion of a CCD.  The pipeline takes the brightest 200 unsaturated sources in each subregion, which are chosen to be unmasked, relatively unblended, and to have positions which did not move by more than 1 pixel in the previous iteration.  The images of these stars, after neighbors have been subtracted according to the best model so far, are used to fit the parameters of the PSF model for the next iteration.

A primary goal of the ideal-seeing PSF is to allow the pipeline to describe diffraction spikes and other features in the wings of the PSF.  These features may not be constant over the focal plane, but treating them as constant is an improvement over neglecting them altogether.  The ideal-seeing PSF is derived by averaging the PSF of many stars in the focal plane over all CCDs and several images on a night with very good seeing.  This PSF is then deconvolved with a Moffat PSF using Richardson-Lucy deconvolution to obtain a PSF with better seeing than obtained on any night.  The wings  of the deconvolved PSF are then fit as the sum of 6 Moffat PSFs plus power law diffraction spikes and horizontal and vertical features.  The resulting noise-free model PSF wings are then blended with the central high $S/N$ core of the deconvolved PSF to produce the final ideal-seeing PSF.

\subsection{Failure modes}
\label{subsec:catfailure}

The \texttt{crowdsource} analysis can fail in a number of ways.  Because \texttt{crowdsource} attempts to explain all flux in the image as due to stars and a smooth background, violations of this assumption and imperfections in the PSF can introduce spurious sources into the catalog.

By far the most common cause of spurious sources is very bright stars.  The \texttt{crowdsource} PSF wings are based on a single ideal-seeing model PSF that is fixed for all images in each band.  This ideal-seeing PSF is then convolved with a Moffat profile, whose parameters may vary linearly in each image.  This limited allowed variation, however, is not adequate to describe all of the features in the wings of the PSF.  For instance, the relative amount of flux in the different diffraction spikes varies over the focal plane and cannot be captured in the \texttt{crowdsource} model.  Accordingly, some spurious sources are present near most bright stars (brighter than roughly 12th mag).  Extremely bright stars, like $\alpha$ Cen, can produce thousands of spurious sources.  Demanding that a source be detected in multiple filters, or at least detected multiple times, can significantly reduce the number of this kind of spurious source.

Bright stars are especially problematic in $Y$ band because of the extremely long PSF features extending vertically and horizontally from the PSF center in $Y$ band.  The \texttt{crowdsource} PSF model describes these features accurately, but for sufficiently bright stars they are highly significant even several hundred pixels from the star's center, beyond the range that the \texttt{crowdsource} PSF extends.  Accordingly, very bright stars in $Y$ band can lead to clusters of spurious sources 150 pixels away in the horizontal and vertical directions, where the PSF model ends and flux is left on the image.  These sources will not be detected in any other bands, but will contaminate the $Y$ band photometry in their vicinity slightly.

Because the \texttt{crowdsource} PSF model is expected to be imperfect, no attempt is made to find sources too near existing, brighter sources.  This means that many close blends are not identified.  Blends are usually not deblended if the two stars involved are separated by less than one half FWHM, though the flux ratio of the blended sources also matters.  More aggressive deblending is possible, but must be balanced against avoiding incorrect deblending of single sources due to PSF-misestimation.

In extremely dense regions like Baade's window, the \texttt{crowdsource} algorithm never completely converges.  A very small fraction of faint but clearly significant sources are never identified, and extremely complicated blends of hundreds of stars are not fully worked out.  It is not clear how much better one can do in these regions, however: they are extremely blended, and the derived model images account for most of the flux.

For simplicity, the \texttt{crowdsource} pipeline adopts a fixed size, $5\arcsec\times5\arcsec$ PSF stamp independent of the seeing of the image.  In the worst seeing conditions, this PSF stamp size does not enable the pipeline to track flux far enough into the PSF wings.  This can lead to spurious sources $2.5\arcsec$ from true sources, but only in images with $\mathrm{FWHM} > 2\arcsec$, roughly 0.4\% of all images.  Most such spurious sources should be detected in only one of the three observations of each part of the sky, facilitating their rejection from scientific analyses.

The DECaPS pipeline aims to characterize all sources in individual images down to a point-source depth of 5$\sigma$.  In practice, DECaPS achieved something closer to 6$\sigma$ in typical images, owing to crosstalk between the sky subtraction routine and the source detection routine.  Sky subtraction was performed by median-filtering the image in a 20$\times$20 pixel region.  Sources in this region will bias the sky high.  Having been detected, the sources are subtracted and the sky is refit.  This process is iterated several times, so the effect on the final fluxes is small.  However, this means that once sources are detected at 5$\sigma$ after an initial sky subtraction has been performed, they are later detected at apparently higher significance once the sky has been lowered.  We could have eliminated this effect by using larger region for the median filtering or considering a local background subtraction when defining the initial source significance, but both of these approaches also have downsides.  Accordingly, the final survey does not quite reach 5$\sigma$.

\section{Photometric Calibration}
\label{sec:cal}

The DECaPS photometric calibration follows the photometric calibration from PS1 \citep{Schlafly:2012, Magnier:2013, Magniercal:2016} closely, which was modeled on the calibration of the SDSS \citep{Padmanabhan:2008}.  The basic idea is to solve for the throughput of the atmosphere and detector as a function of time, airmass, and location of the observed source in the focal plane, in order to minimize the variance in repeated measurements of the same sources.  Since DECaPS was designed to observe most stars in its survey area three times, and since billions of stars were observed, precise measurements of the system throughput and its variation can be made.

In PS1, we adopted a simple model for the system throughput: a single constant throughput on each night (i.e., a system zero point), an airmass term on each night, and a constant-in-time flat field correction.  In DECaPS, the same model was used, with an additional simplification: the airmass term was held constant in each filter over all nights of the survey.  This simplification was necessary because of the small range of airmass observed on each night, coupled with imperfect aperture corrections (see \textsection\ref{subsec:callimitations}).  Each filter is calibrated independently.  The flat field correction treats each $256\times256$ pixel portion of each detector independently.

\begin{figure}[htb]
\begin{center}\includegraphics[width=\columnwidth]{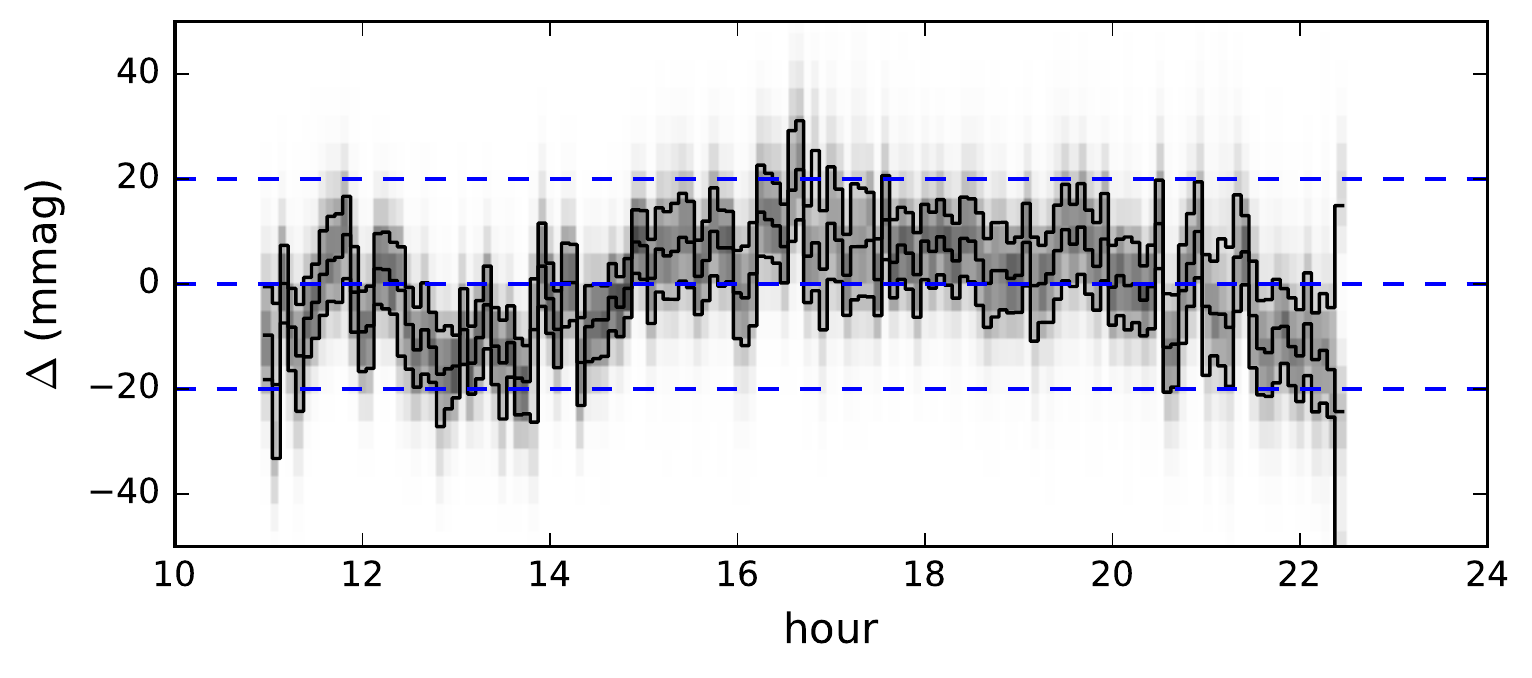}\end{center}
\caption[Photometry calibration]{
\label{fig:qacal}
The distribution of residuals as a function of hours since noon UTC on the night of April 29, 2017.  Each column shows the distribution of residuals at a particular time, relative to the average magnitude of all detections of those sources.  Solid lines show the 16, 50, and $84^\mathrm{th}$ percentile of the residuals.  At any given time, the residuals are consistent to better than about 8~mmag, and overall zero points are predicted to better than 1\%.  The dominant systematic in the overall zero points is the aperture correction of the PSF.
}
\end{figure}

The performance of this simple model is excellent.  Table~\ref{tab:calperformance} gives the accuracy of the calibration model in predicting the zero points of individual images, as compared to a model where the zero point is allowed to float in each image ($\sigma_\mathrm{ZP}$), and the average scatter in differences in fluxes of bright stars on an image from their mean fluxes over all observations ($\sigma_\mathrm{bright}$).  On a typical image, bright stars' fluxes agreed with the mean fluxes over all observations of those stars to better than 8 mmag in all filters.  $Y$ band showed the best performance, with 6 mmag scatter.  This agreement is superior to the results we achieved in PS1, where 15 mmag scatter in the flux of bright stars was typical.  Figure~\ref{fig:qacal} shows the residuals of the calibration on a typical night of the survey.

\begin{deluxetable}{cll}
\tablewidth{\columnwidth}
\tablecaption{Photometric Calibration Performance}
\tablehead{
\colhead{Filter} & \colhead{$\sigma_\mathrm{ZP}$ (mmag)} & \colhead{$\sigma_\mathrm{bright}$ (mmag)}
\label{tab:calperformance}
}
\startdata
g &   10.3 &   7.6 \\
r &   10.2 &   6.9 \\
i &   8.6  &   6.7 \\
z &   9.4  &   6.4 \\
Y &   8.3  &   6.0 \\
\enddata
\tablecomments{
The performance of the photometric calibration.  The simple nightly zero point and airmass term described the zero points of individual images to better than about 1\% ($\sigma_{ZP}$).  Bright stars on each image agreed with the mean fluxes of those stars over all observations with a scatter of better than 8 mmag in each band ($\sigma_\mathrm{bright}$).
}
\end{deluxetable}

\begin{figure}[htb]
\begin{center}\includegraphics[height=0.88\textheight]{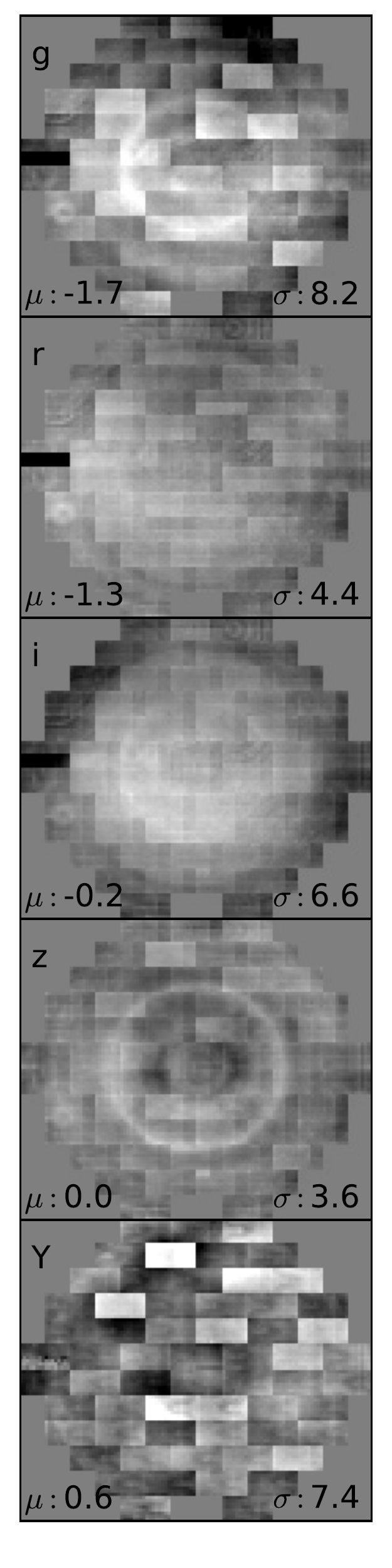}\end{center}
\caption[Flat field correction]{
\label{fig:flat}
The DECaPS flat field correction in the $grizY$ bands.  The corrections are in all cases small, with typical values of about 6 mmag.  The features that are detected, however, are detected with extremely high significance, and relate to real features in the DECam CCDs and their processing.  The mean and standard deviation of the flat field correction are given in the lower left and right of each panel, respectively.
}
\end{figure}

The flat field correction derived in the calibration is shown in Figure~\ref{fig:flat}.  The flat fields show a large number of features.  Most prominent is amplifier B of CCD S7, shown as a saturated black region at the left of the $gri$ flat field correction maps.  No photometry from S7 amplifier B is used in the mean magnitudes we compute for DECaPS sources.  Potentially, photometry from this amplifier could have been separately calibrated, but this was not implemented for DECaPS.

The $g$ and $Y$ band flat fields feature significant CCD-to-CCD offsets.  This may be caused by the combination of varying throughput as a function of wavelength between the different DECam CCDs, and the significantly redder typical spectrum of DECaPS sources than the spectrum used to determine the DECam flat field.  The strong radial gradient in $i$ is thought to be due to variation in the $i$ filter bandpass as a function of angle.  The $z$ band shows a significant ``pupil ghost'' pattern, which is related to the technique used to determine the initial DECam sky flats; to a lesser extent, $g$ and $i$ show similar effects.  The $<0.01$ mag $z$-band pattern is a small residual remaining after the much larger pupil corrections made in the CP.  A hint of tree rings \citep{Plazas:2014} is noticeable in the upper right CCD in $r$ and $i$ bands.  The DECaPS flat field does not have the resolution to fully resolve the tree rings, and a couple of mmag leak into the flat field corrections.  In any case, the CP flat fields are quite good: the flat field corrections have less than 9 mmag scatter in all filters, and the correction for the worst filter, $g$, is largely caused by varying CCD throughputs with wavelength, which no single flat field could correct.

\subsection{Limitations of the Photometric Calibration}
\label{subsec:callimitations}

The calibration model could be improved in several respects.  It ignores ``systematic chromatic errors'' \citep{Li:2016}, essentially neglecting the fact that changes in system throughput have a wavelength dependence.  This causes stars to require different corrections into a ``standard'' system according to their spectra, but the DECaPS calibration performs the same correction for all objects.  More physics-based analyses can readily accommodate these effects, and are already being performed for DECam \citep[e.g.][]{Burke:2017}. 

The photometric calibration model is only able to predict the zero point of a given exposure to within about 9 mmag ($\sigma_\mathrm{ZP}$).  While not bad, this is substantially worse than achieved by PS1, which had 3 mmag precision.  Evidently the simple single nightly system zero point and airmass term provided a much better description of the PS1 images' zero points than the similar model adopted for DECaPS.  The dominant unmodeled source of variation in the DECaPS images' zero points turns out to be the ``aperture correction'' of the crowded DECaPS photometry.  If the pipeline's PSF were perfectly accurate, then the measured PSF flux should equal the true flux on average.  With an imperfect PSF, the measured PSF fluxes become systematically biased.  The bias can be addressed by measuring aperture magnitudes for bright stars in very large apertures, and comparing them with the measured PSF fluxes of those stars.  The difference is related to the aperture correction and can be applied to all PSF fluxes to correct them to total fluxes.  This procedure is challenging in crowded fields, so DECaPS did not perform an aperture correction.  Comparison of different PSF fitting routines indicates that the pipeline was too inflexible in fitting the PSF roughly $2\arcsec$ from the center of the PSF, leading variation in the shape of the PSF at this radius to cause apparent $\sim 7$ mmag variations in measured PSF flux---that would have been absorbed by the aperture correction, had one been made.  Future reductions of the DECaPS imaging can improve on these results by using a more flexible PSF model, or alternatively \texttt{crowdsource} could be adapted to compute an aperture correction on neighbor-subtracted images.

Other improvements on the DECaPS photometric calibration are possible.  We neglect in this photometric calibration the ``tree ring'' pixel-area effects discussed in \citet{Plazas:2014}, which are expected to contribute a few mmag of photometric noise.  Eliminating systematic chromatic errors is difficult without knowing the spectral energy distributions of the stars within the filter bandpasses, but upcoming Gaia low-resolution spectrophotometry will eliminate this problem for bright stars.  By addressing these shortcomings, we expect that significantly better than 5~mmag photometry for typical stars is possible over large areas with DECam, even in crowded regions.

\subsection{Absolute Zero Point}
\label{subsec:abszp}
The absolute zero point was obtained by reference to PS1, in combination with predicted colors in the DECam bands.  The procedure is laid out in \citet{Scolnic:2015}: stellar energy distributions are taken from calibrated HST measurements and integrated over the PS1 and DECam filter bandpasses to predict the color differences between the surveys.  Absolutely calibrated PS1 photometry is then transformed to absolutely calibrated DECam photometry using the derived color transformations for blue stars ($0.25 < g_\mathrm{PS1} - i_\mathrm{PS1} < 1$) where the transformations are most accurate.  The resulting absolutely calibrated, synthesized DECam photometry is then used to fix the absolute zero points for DECaPS.  Ultimately, this calibration derives from Hubble Space Telescope measurements from \citet{Bohlin:2014}.  A calibration field at $(\alpha, \delta) = (100\degree, -27\degree)$ is used to link the two surveys together.

The DECaPS absolute calibration is then no better than the PS1 absolute calibration.  DECaPS adopted the original PS1 absolute calibration from \citet{JTphoto}; this has since been superseded by the calibration of \citet{Scolnic:2015}.  To shift DECaPS to the absolute calibration of \citet{Scolnic:2015}, offsets of 0.020, 0.033, 0.024, 0.028, 0.011  mag must be added to the DECaPS $grizY$ magnitudes, respectively.

\section{Survey Characteristics}
\label{sec:characteristics}

\subsection{Photometry}
\begin{figure*}[htb]
\begin{center}\includegraphics[width=\textwidth]{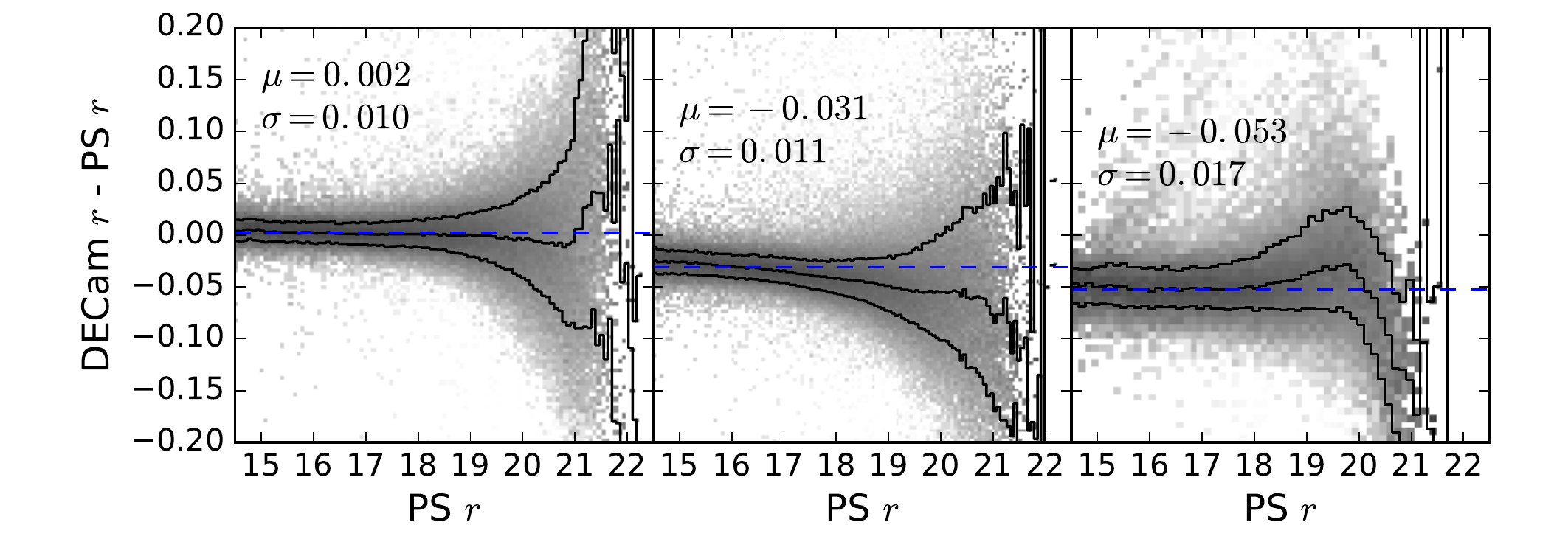}\end{center}
\caption[PS Comparison]{
\label{fig:pscomp}
Comparison between DECaPS and PS1 photometry in the $r$ band.  Other bands are similar.  The left panel shows a calibration field at $(l, b) = (236\degree, -14\degree)$; the middle panel shows a field at $(l, b) = (115\degree, 0\degree)$; and the right panel shows a field at $(l, b) = (0\degree, 0\degree)$.  In general, agreement is good, though there is a 0.05 mag zero point offset between the Galactic center field and the calibration field.  This may be due to the color correction, which was developed in a field of little reddening, while the Galactic center field (and, to a lesser extent, the other midplane field) is heavily reddened.  The mean photometric offset and the scatter between the magnitudes in the two surveys for stars greater than 19th mag is shown; typically, the two surveys agree to better than 2\%.  
}
\end{figure*}

The accuracy of DECaPS photometry can be tested by comparison with PS1 over the limited area of overlap.  The DECaPS survey extends slightly beyond $\delta = -30\degree$ toward the Galactic center and anticenter to enable this comparison, and also contains a few calibration fields off the plane above $\delta = -30\degree$.  The results of this comparison for the $r$ band are shown in Figure~\ref{fig:pscomp}.  The first panel shows the comparison on a calibration field at $(l, b) = (236\degree, -14\degree)$, the second panel shows the comparison at $(l, b) = (115\degree, 0\degree)$, and the third panel shows the comparison at $(l, b) = (0\degree, 0\degree)$.

The two surveys show good agreement in their photometry; the bright-end root-mean-square (RMS) difference is around 0.02 mag, and smallest in the unreddened calibration field.  There is a modest trend in the photometry with magnitude in the anticentral field (0.03 mag over from 15th to 21st mag).  The trend may be again connected to reddening (fainter stars are systematically more reddened), but we see the opposite trend, if any, in the most reddened Galactic center field.  No two fields have quite the same overall zero point; the DECam photometry becomes systematically brighter relative to the PS1 photometry in the more reddened fields.  The Galactic center field is offset by $-0.05$ mag relative to the calibration field.  We attribute this to the imperfect color transformations in the presence of reddening, and conclude that the calibration is more uniform than about 0.05 mag, though expectations from the internal repeatability suggest that the calibration should be significantly better than this.  Other filters show qualitatively similar trends.

\begin{figure}[htb]
\begin{center}\includegraphics[width=\columnwidth]{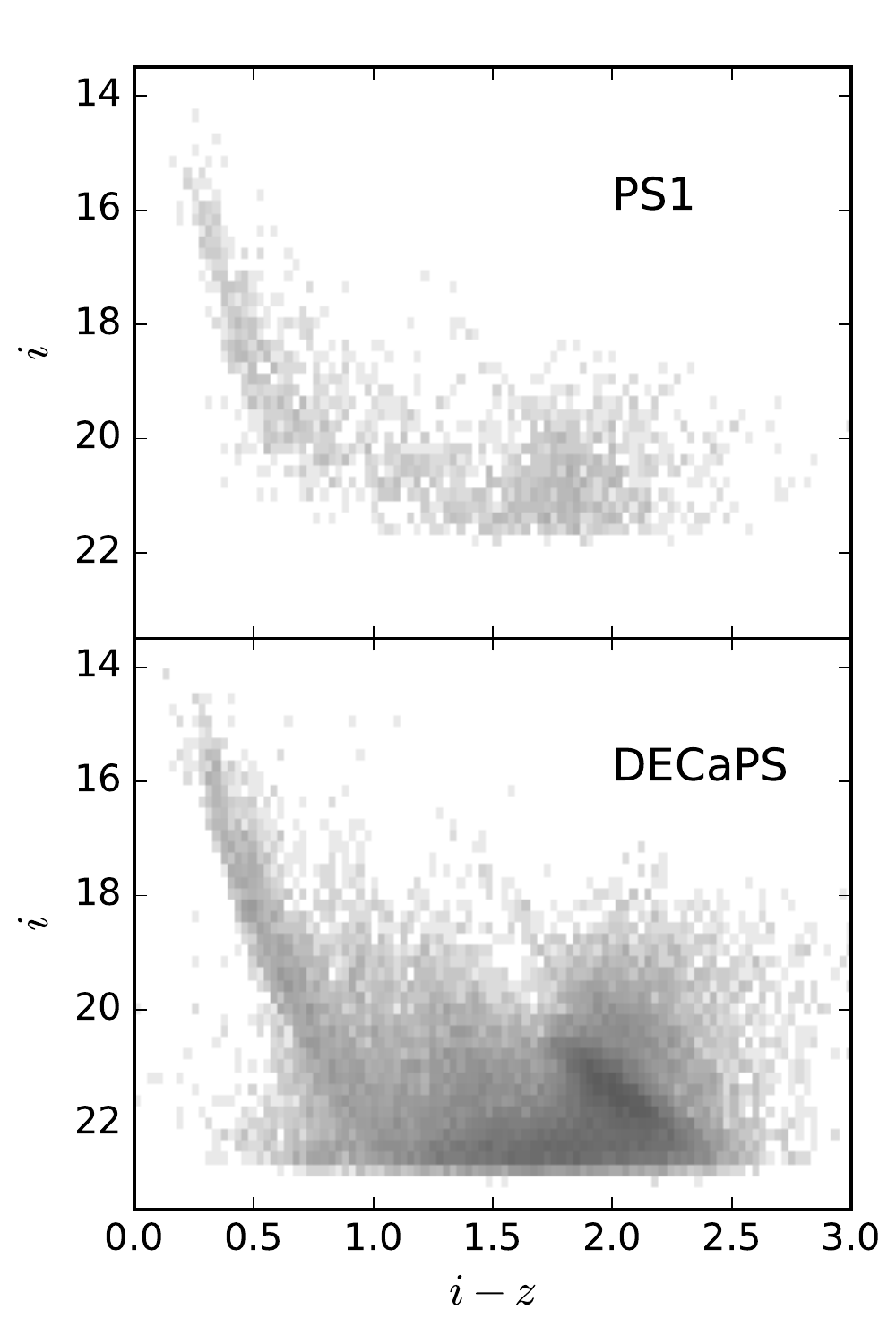}\end{center}
\caption[PS CMD Comparison]{
\label{fig:pscmdcompare}
Comparison between DECaPS and PS1 $i-z$, $z$ color-magnitude diagrams, for a $0.1\degree$ radius beam around around $(l, b) = (0.3, 0.7)$.  The region is modestly crowded.  The DECaPS CMD is significantly deeper and clearly shows the red clump in the bulge, which is hard to separate in the PS1 diagram.
}
\end{figure}

Another comparison between PS1 and DECaPS is shown in Figure~\ref{fig:pscmdcompare}.  The superior depth and deblending of the DECaPS pipeline makes for a much more obvious bulge red clump at $i=21.5$, $i-z = 2.0$; the DECaPS measurements clearly improve on the PS1 measurements in the small overlap region between the two surveys.  On the other hand, a few blue sources around $i=22.0$, $i-z = 0.5$ are presumably largely spurious deblending errors in DECaPS; they are absent from the PS1 CMD, and are largely removed by a modest cut on the blendedness of a detection (\texttt{fracflux}, see Table~\ref{tab:mergedcatalog}).

\subsection{Color transformations}

PS1 photometry is of qualitatively similar depth to DECaPS photometry.  The PS1 filter bandpasses are moreover very similar to the DECaPS bandpasses.  Accordingly, it is useful to have a set of color transformations to allow comparison between DECam $grizY$ filters used in DECaPS and the PS1 $grizy$ filters.  To develop these color transformations and improve the stability of the photometric calibration, DECaPS observed a calibration field at $(\alpha, \delta) = (100\degree, -27\degree)$ on many different nights of the survey.  This field has low reddening ($\approx 0.15$ mag $E(B-V)$) but is at low latitude ($b = -14\degree$), meaning that it contains many stars suitable for deriving color transformations.

\begin{figure*}[htb]
\begin{center}\includegraphics[width=\textwidth]{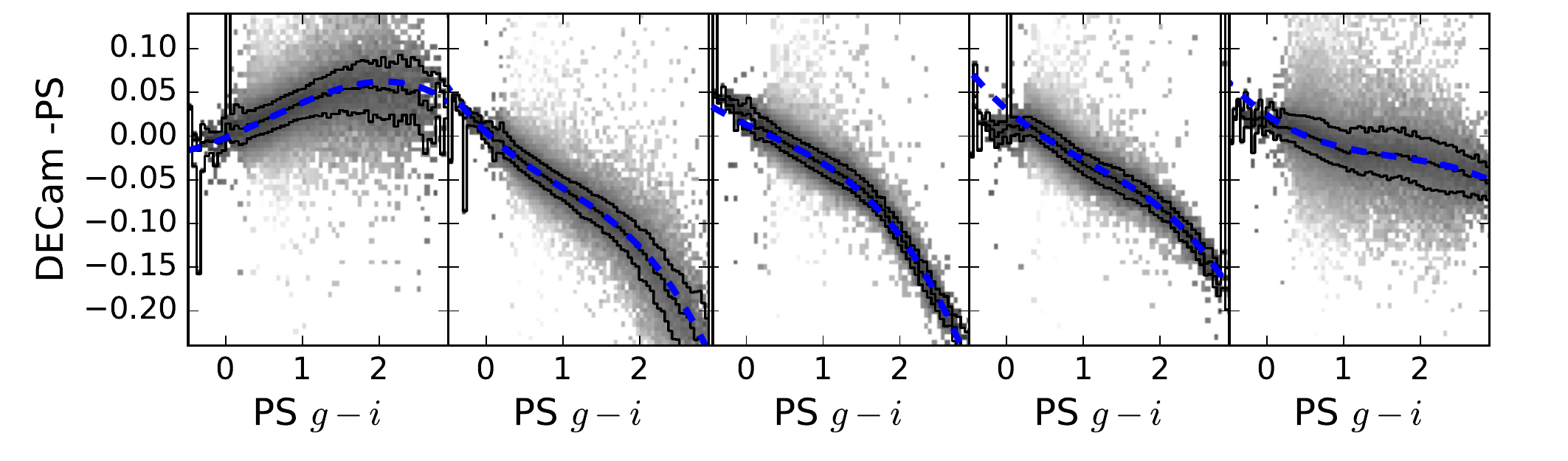}\end{center}
\caption[Color Transformations]{
\label{fig:cterms}
Color transformations between DECam and PS1.  The difference between DECam and PS1 photometry in the $grizY$ bands is shown as a function of PS1 $g-i$ color, for bright stars in the vicinity of $(\alpha, \delta) = (100\degree, -27\degree)$.  A cubic fit to the difference is shown by the blue dashed lines; these are the color transformations adopted in DECaPS.  Photometric agreement is good, about 0.015 mag.
}
\end{figure*}

Color transformations between DECam and PS1 were fit with a cubic polynomial in the PS1 $g-i$ color.  The color transformations are given in Equation~\ref{eq:cterms}.  Only point sources brighter than 19th mag in a nearby filter that was neither $g$, $i$, nor the filter of interest were used to determine the transformations.  Furthermore, only stars with $0 < g_\mathrm{PS1}-i_\mathrm{PS1} < 2.9$ were used; the color transformation outside this broad range requires more parameters and more stars to constrain.  Figure~\ref{fig:cterms} shows the derived color transformations.  After color correction, the RMS difference between DECaPS and PS1 photometry is about 0.015 mag.
\begin{widetext}
\begin{align}
\label{eq:cterms}
c &= g_\mathrm{PS1}-i_\mathrm{PS1} \\
g_\mathrm{DECam} - g_\mathrm{PS1} &= 0.00062 +  0.03604 c   +0.01028 c^2 -  0.00613 c^3 \\
r_\mathrm{DECam} - r_\mathrm{PS1} &= 0.00495 -  0.08435 c   +0.03222 c^2 -  0.01140 c^3 \\
i_\mathrm{DECam} - i_\mathrm{PS1} &= 0.00904 -  0.04171 c   +0.00566 c^2 -  0.00829 c^3 \\
z_\mathrm{DECam} - z_\mathrm{PS1} &= 0.02583 -  0.07690 c   +0.02824 c^2 -  0.00898 c^3 \\
Y_\mathrm{DECam} - y_\mathrm{PS1} &= 0.02332 -  0.05992 c   +0.02840 c^2 -  0.00572 c^3
\end{align}
\end{widetext}

\subsection{Sample Color Magnitude Diagrams}

The DECaPS footprint includes open clusters, globular clusters, nearby molecular clouds, distant molecular clouds, and regions with a wide range of stellar densities.  Accordingly, color magnitude diagrams from DECaPS show rich diversity.

\begin{figure}[htb]
\begin{center}\includegraphics[width=\columnwidth]{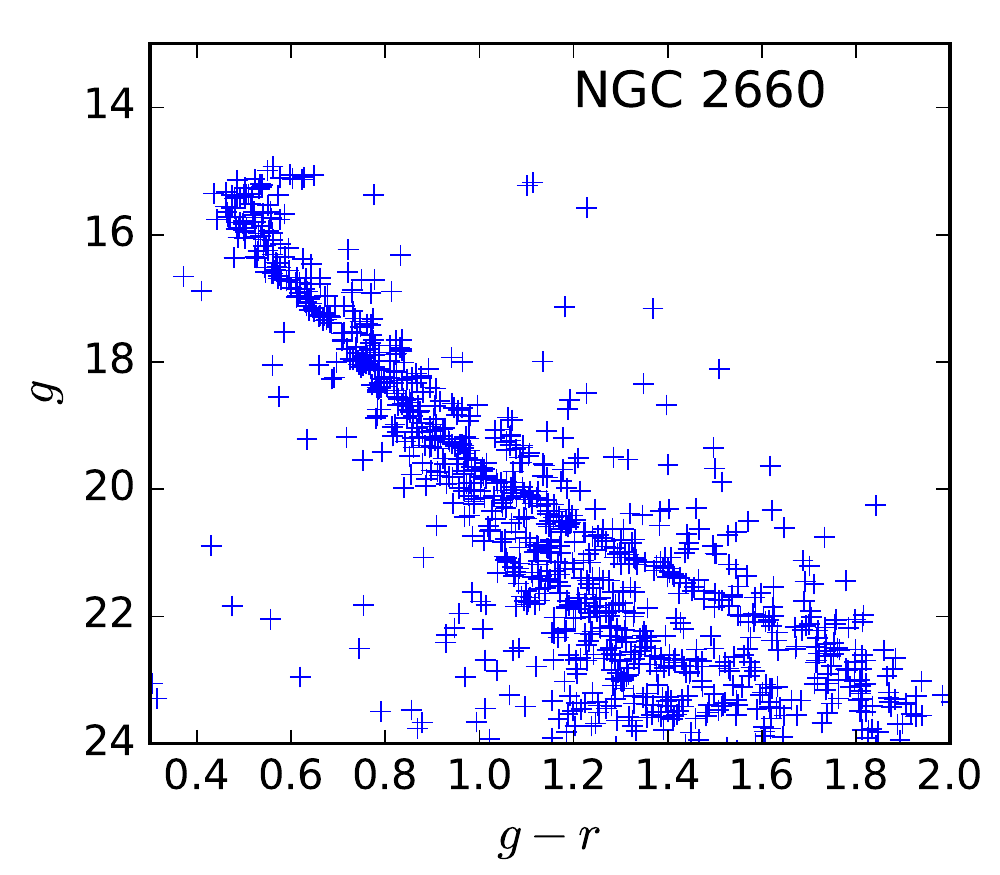}\end{center}
\caption[NGC 2660]{
\label{fig:ngc2660cmd}
The DECaPS $g$, $g-r$ color-magnitude diagram of NGC 2660.  A sharp main sequence is superimposed on a background of field stars whose significant and increasing reddening with distance causes them to overlap significantly with the NGC 2660 sequence.  A second sequence roughly 0.7~mag above the NGC 2660 sequence is also visible.
}
\end{figure}

To demonstrate the precision of the DECaPS photometry, Figure~\ref{fig:ngc2660cmd} shows the $g$, $g-r$ color-magnitude diagram of stars within 0.03\degree\ of the center of NGC 2660, an approximately 1 Gyr-old open cluster at a distance of 2.7 kpc \citep{Sandrelli:1999}.  Reddening of the field stars in the region leads them to have a very similar CMD to that of NGC 2660, but nevertheless the sharp NGC 2660 sequence clearly stands out from the background.  A secondary sequence roughly 0.7~mag brighter than the primary sequence may indicate a substantial fraction of binaries \citep{Sandrelli:1999}.  

\begin{figure*}[htb]
\begin{center}\includegraphics[width=\textwidth]{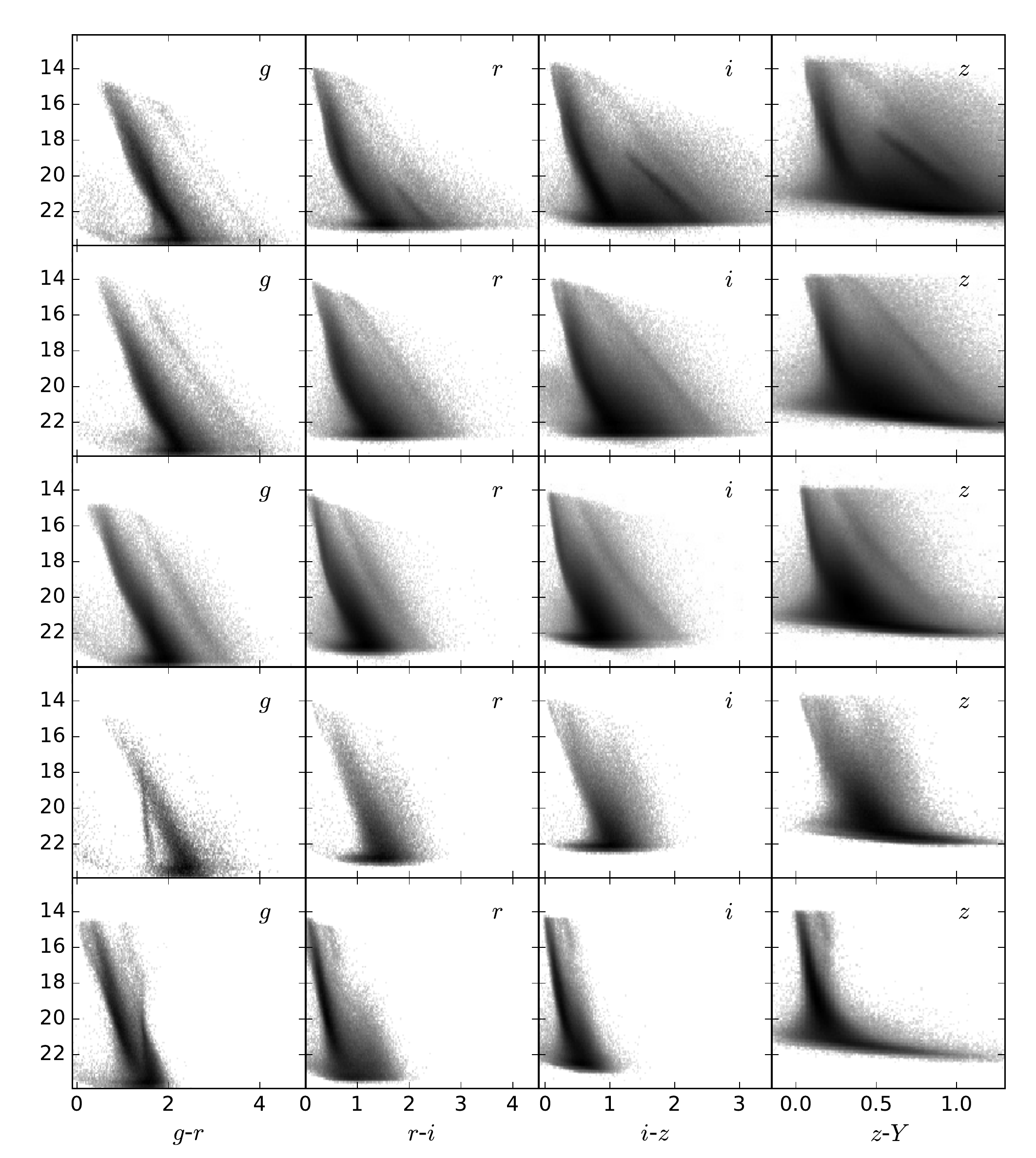}\end{center}
\caption[CMDs]{
\label{fig:cmds}
DECaPS color-magnitude diagrams of 1\degree\ radius beams in the Galactic plane.  Each row shows a different field in the Galactic midplane; from top to bottom, the rows correspond to $l=0\degree, -30\degree, -60\degree, -90\degree$ and $-120\degree$.  The diagrams show a diverse array of stellar populations and extensive reddening.}
\end{figure*}

The NGC 2660 color-magnitude diagram demonstrates the value of DECaPS photometry to constrain stellar populations in the Milky Way.  However, typical color-magnitude diagrams in the footprint are much more complicated than that in the vicinity of NGC 2660.  Figure~\ref{fig:cmds} shows a selection of CMDs in different filters and Galactic longitudes from DECaPS.

Each row of Figure~\ref{fig:cmds} shows the stars in a one degree radius beam in the Galactic plane ($b=0\degree$) at different Galactic latitudes, stepping out in 30\degree\ increments from $l=0\degree$ to $l=-120\degree$.  Each diagram shows a strong sequence of blue stars that are reddened as one proceeds to fainter magnitudes: the main sequence and main-sequence turn-off in the presence of increasing extinction along the line of sight.  The Galactic center field (top row) additionally features a fainter, redder sequence: the red clump in the Galactic bulge.  The outer Galaxy fields show relatively unreddened nearby stars (e.g., the unreddened M-dwarf sequence with $g-r = 1.5$ mag at $l=-90\degree$), on top of fainter, redder populations.

\subsection{Photometric Uniformity}

\begin{figure*}[htb]
\begin{center}\includegraphics[width=\textwidth]{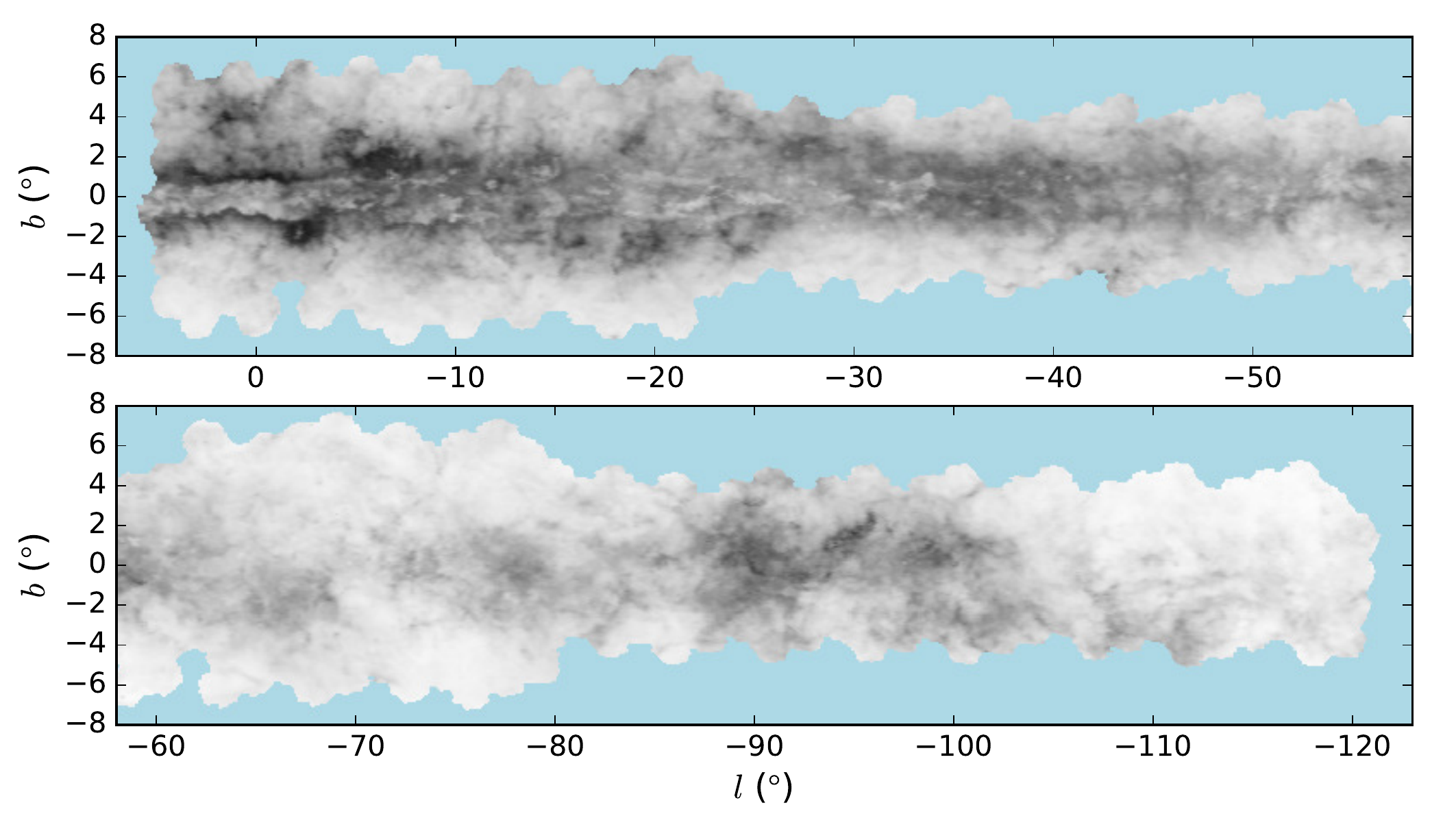}\end{center}
\caption[Median $r-i$ color]{
\label{fig:medricol}
The median $r-i$ color of stars detected in DECaPS with $Y$ magnitude brighter than 18.  The $r-i$ color is dominated by reddening from dust and its distribution along the line of sight; major dust clouds like the Vela Molecular Ridge at $l=-90\degree$ are prominent.  There are no clear signs of errors in the photometric calibration.  The discontinuity in $r-i$ color at $|b| = \pm 1\degree$, $|l| < 5\degree$ is the result of increasingly dense molecular clouds leading the median star to be in the foreground of these clouds, rather than in the background.}
\end{figure*}

The internal consistency obtained in the DECaPS photometric calibration (\textsection\ref{sec:cal}) suggests that systematics in the photometric calibration over the footprint should be about 0.01 mag in size.  This estimate can be tested by measuring the consistency of stellar colors over the DECaPS footprint \citep[e.g.][]{High:2009, Schlafly:2010}.  Unfortunately, because the DECaPS footprint features very large extinction, the typical colors of stars are completely dominated by reddening.  Nevertheless, Figure~\ref{fig:medricol} shows the median $r-i$ color of stars with $Y$ band magnitudes brighter than 18 over the footprint.  White to black spans 0.2 to 2.5 magnitudes.

The Figure is clearly dominated by extinction in the Galactic plane.  Well known molecular clouds like the Vela Molecular Ridge around $l = -90\degree$ and the Pipe Nebula at $(l, b) = (1\degree, 4\degree)$ are apparent.  In the very inner Galaxy ($|l| < 5\degree$), the median $r-i$ color of stars gets steadily redder as $b\rightarrow 0\degree$, until $b \sim 1\degree$, where the $r-i$ color suddenly becomes more blue.  This occurs because behind dense molecular clouds, few stars are detected, leading the median-colored star to become a star in the foreground of those molecular clouds, rather than a star in their background.  No clear signs of photometric errors due to the photometric calibration are present in Figure~\ref{fig:medricol}, but given the wide range of colors shown and the large variations in color due to dust reddening, this test does not strongly constrain the quality of the photometric calibration.

\subsection{Astrometry}
Precise astrometry was not a major objective of DECaPS.  The \texttt{crowdsource} pipeline computes best fit $x$ and $y$ positions for the sources it detects, and transforms these into celestial coordinates using the CP-computed World Coordinate System (WCS) with no additional steps or calibration.  

This is a simple but problematic approach.  The original CP WCS may have a different notion for the ``center'' of a source than the \texttt{crowdsource} pipeline.  Ultimately, both pipelines attempt to use the centroids of the PSF as the center, but the centroid may have varying definitions depending on how much weight is put on flux far out in the wings of the point spread function.  This is a small effect, but would motivate directly calibrating the \texttt{crowdsource} positions to external catalogs rather than using a separate set of centroids in the calibration.

Additionally, the CP WCS calibration changed over the course of DECaPS.  Starting with CP version 4, world coordinate systems were computed using Gaia as the astrometric reference catalog, while before that point 2MASS was used.  This is unfortunate for DECaPS, because it means that the astrometry from the first part of the survey is slightly inconsistent with the last part of the survey.  Still, the 2MASS and Gaia astrometry are very similar, and DECaPS only requires that the astrometry be similar enough to one another so that most stars can be merged with an association distance of 0.5\arcsec.  However, other applications of the DECaPS observations, like the construction of coadded stacks, may require greater fidelity.

\begin{figure*}[htb]
\begin{center}\includegraphics[width=\textwidth]{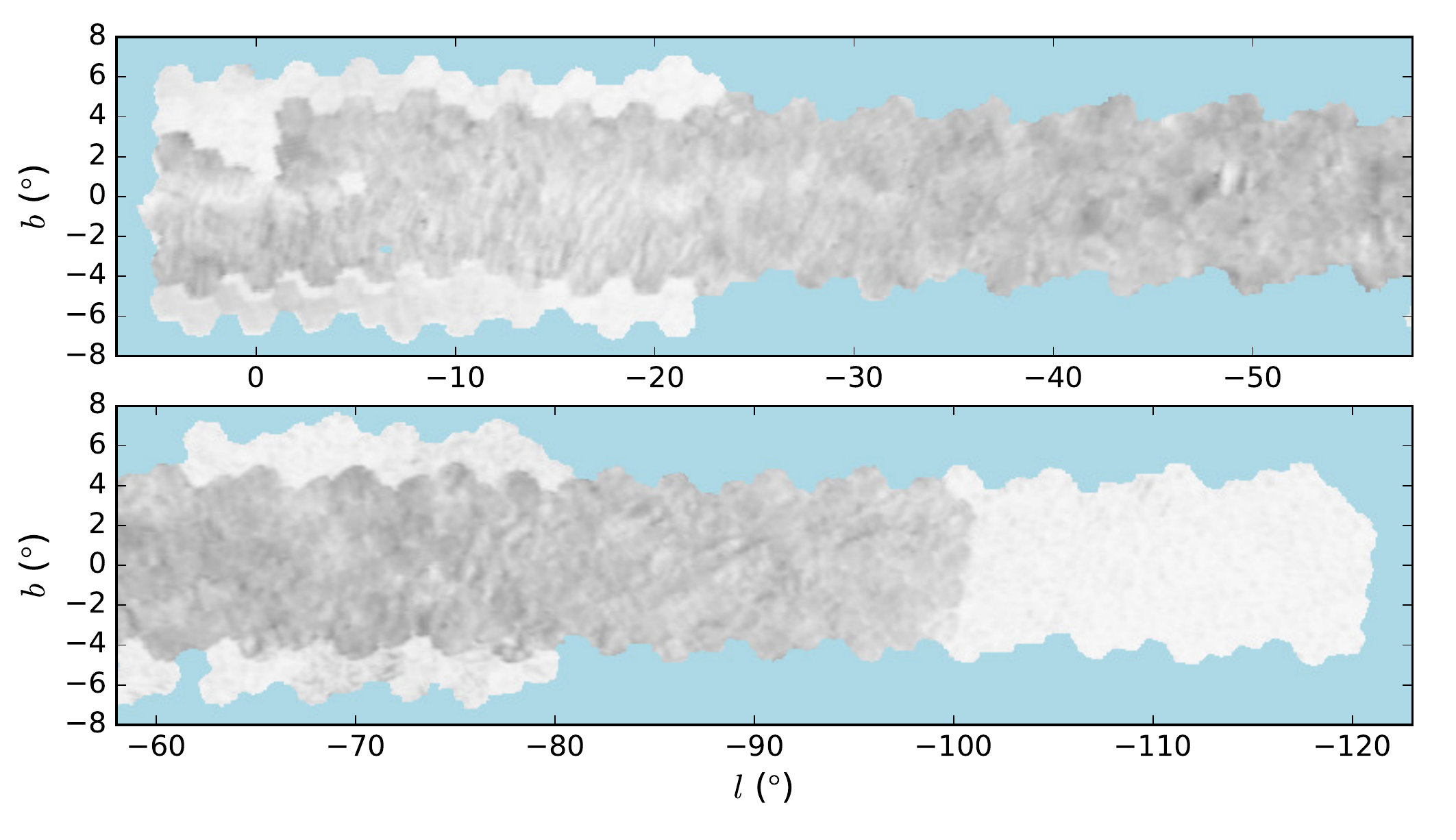}\end{center}
\caption[Mean position difference between DECaPS and Gaia]{
\label{fig:gaiacomp}
The mean difference in position between DECaPS and Gaia as a function of position over the DECaPS footprint.  White to black spans 0.0--0.5\arcsec.  Regions of the footprint which were processed only using CP version 4 or later have typical differences close to zero, while regions processed using earlier versions of the CP have typical differences of 0.1\arcsec.}
\end{figure*}

Figure~\ref{fig:gaiacomp} shows the comparison between DECaPS and Gaia astrometry for stars brighter than 19th mag in $r$.  The mean difference in position between the DECaPS stars in each part of the sky is shown; white to black is 0--0.5\arcsec.  Typical values are 0.1\arcsec\ in the part of the footprint that was primarily observed using CP versions earlier than 4, while the CP 4-only fraction of the footprint shows much lower residuals.  This level of difference between DECaPS and Gaia is unproblematic from a DECaPS perspective; the astrometry is much better than the 0.5\arcsec\ necessary to merge the catalogs from the individual images.

\section{Data Release}
\label{sec:release}
All DECaPS images and catalogs are now publicly available.  Raw and CP processed images are available through the NOAO Archive (NOAO Prop. ID 2014A-0429, 2016A-0327, and 2016B-0279).  Catalogs, both for individual images and merged, are available from the survey website, \texttt{http://decaps.skymaps.info}.  The catalogs contain more than twenty billion detections of more than two billion objects.

\subsection{Viewer}
\begin{figure}[htb]
\begin{center}\includegraphics[width=\columnwidth]{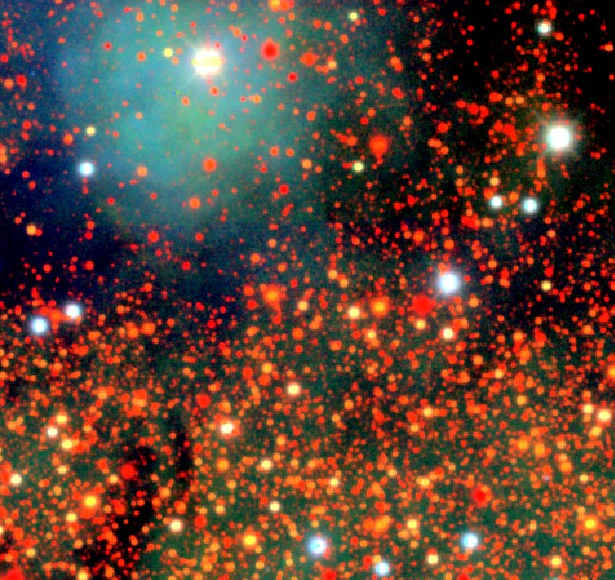}\end{center}
\caption[Viewer: NGC 6188]{
\label{fig:viewer}
The DECam Legacy Survey Viewer view of a small part of NGC 6188.  The three-color image is made of of DECaPS $g$, $r$, and $z$ stacked images. 
}
\end{figure}

DECaPS images have been imported into the DECam Legacy Survey viewer \citep{DESI:2016a, DESI:2016b}.  Zero points from the photometric calibration (see \textsection\ref{sec:cal}) are applied to each image, and a single constant sky is removed, according to the median sky estimate of the \texttt{crowdsource} analysis.  Images are then projected onto a common frame and coadded with inverse variance weighting to produce stacks.  Stacks in the $g$, $r$, and $z$ bands are then combined to produced color images and made available online in a zoomable and pannable viewer.  Figure~\ref{fig:viewer} shows an example image from the viewer.

Dramatic dust clouds are evident in the top half of the image, where nearly all stars are extremely red due to foreground extinction.  A reflection nebula around one bright star lights up a large region with relatively blue reflected light. Several foreground stars are immediately identifiable from their bright, white colors, as compared with the red background stars.  The greenish continuum emission at the bottom of the image is H$\alpha$ and reflected light.

In order to preserve many of these continuum features in the viewer, a single background value is subtracted from each DECam CCD before coaddition.  This does a good job at preserving features with spatial scales much smaller than a CCD, but corrupts structures with a larger spatial scale.  Accordingly, some of the brightest and largest objects in the sky get corrupted in their display in the viewer.  For instance, near $\alpha$~Centauri and $\eta$~Carinae, the per-CCD sky subtraction removes a significant amount of flux from the objects themselves and leads to a multicolored checkerboard pattern around them in the viewer.  A more advanced sky subtraction routine that matches the sky in overlapping images would solve this problem.  However, since this problem only affects a small number of very bright objects, and since stellar photometry is not affected, this issue has not been addressed.

The viewer contains analogous stacks for the \texttt{crowdsource} model images, and the resulting residual images.  These allow for rapid comparison between the data and the model, to assess the reliability of the modeling in any region of the sky.  In general, the model is an excellent description of the data.

\subsection{Catalogs}
\label{sec:catalogs}

DECaPS provides catalogs from individual images analyzed by \texttt{crowdsource} as well as derived, band-merged catalogs giving average photometry in each band for each object.

\subsubsection{Individual Image Catalogs}
\label{subsubsec:individualcatalogs}

Catalogs generated by \texttt{crowdsource} for each invididual DECaPS image are available from the DECaPS website.  These catalogs contain the position of each source, its flux, and the associated uncertainties.  A few other entries assess the reliability of the detection, giving the $\chi^2$ of the detection, and any flags at the location of the central pixel of this source in the CP data quality image.  A ``quality factor'' is also computed; it is the PSF-weighted fraction of pixels that have non-zero weight.  A detection on the edge of a detector would have a quality factor of 0.5, while a good detection has a quality factor close to 1.  Blending is assessed via \texttt{fracflux}, the fraction of the flux in the image at the location of the source contributed by the source itself, weighted by the PSF of the source.  This is low for highly blended objects.  Additional fields give further metadata: the FWHM of the PSF at the location of each source, and the value of the sky and gain.  These fields, and short descriptions, are provided in Table~\ref{tab:individualcatalog}.  A total of slightly over 20 billion records are included in the individual image catalogs.

\begin{deluxetable}{ll}
\tablewidth{\columnwidth}
\tablecaption{Individual Image Catalog Fields}
\tablehead{
\colhead{Name} & \colhead{Description}}
\startdata
\label{tab:individualcatalog}
\texttt{x} & $x$ coordinate (pix) \\
\texttt{y} & $y$ coordinate (pix) \\
\texttt{flux} & instrumental flux (ADU) \\
\texttt{dx} & $x$ uncertainty \\
\texttt{dy} & $y$ uncertainty \\
\texttt{dflux} & \texttt{flux} uncertainty \\
\texttt{fluxlbs} & local-background-subtracted flux (ADU) \\
\texttt{dfluxlbs} & \texttt{fluxlbs} uncertainty \\
\hline
\texttt{qf} & PSF-weighted fraction of good pixels \\
\texttt{rchi2} & PSF-weighted fraction average $\chi^2$ \\
\texttt{fracflux} & PSF-weighted fraction of flux from this source \\
\hline
\texttt{flags} & CP flag value at central pixel \\
\texttt{fwhm} & FWHM of PSF at source location (pix) \\
\texttt{sky} & sky at source location (ADU) \\
\texttt{gain} & gain at source location (e-/ADU) \\
\enddata
\tablecomments{
Fields in the DECaPS individual catalogs.  A more complete description is available at the survey web site.  All fields are stored as 32-bit floating point numbers, except for \texttt{x} and \texttt{y}, which are stored as 64-bit floating point numbers, and \texttt{flags}, which is stored as a 32-bit integer.
}
\end{deluxetable}

\begin{deluxetable}{ccc}
\tablewidth{\columnwidth}
\tablecaption{DECaPS Flags}
\tablehead{
\colhead{Bit} & \colhead{Meaning} & \colhead{Exclude?}
\label{tab:flags}
}
\startdata
1 & Bad pixel & Y \\
3 & Saturated & Y \\
4 & Bleed trail & Y \\
5 & Cosmic ray & Y \\
6 & Low weight & Y \\
8 & Long streak & Y \\
\hline
20 & Additional bad pixel & Y \\
21 & Nebulosity & N \\
22 & S7 amplifier B & Y \\
\enddata
\tablecomments{
Flag bits used in DECaPS.  Flag bits 1--8 are inherited directly from the DECam CP.  Flag bits 20--22 are added by DECaPS.  All flags prevent affected sources from having their fluxes included in the object average fluxes, except for the nebulosity flag (21).
}
\end{deluxetable}

The catalogs contain flux estimates and ``local-background-subtracted'' flux estimates, where an additional constant background was left free to vary around each source.  In general, there is little difference between the default fluxes and the local-background-subtracted fluxes, but substantial differences between the two can indicate problems with the photometry.

\subsubsection{Merged Catalogs}

DECaPS also provides merged catalogs that join all detections of a single object on individual images into a single record.  Photometry is then averaged by band, and an overall average position is measured.  The resulting merged catalogs contain the best estimates of the average flux of the object.

The merge is performed by considering exposures one by one.  All detections in each image are compared with the list of currently existing objects and matched at a $0.5\arcsec$ radius.  Matching detections are considered new detections of existing objects, and non-matching detections are considered the first detections of new objects.

With an assignment of individual detections to objects, the mean photometry of each object in the different bands is computed.  The fluxes of the different detections are calibrated using the throughput model of \textsection\ref{sec:cal}.  DECaPS then computes mean and median calibrated fluxes in each band for each object.  Mean fluxes are weighted by the pipeline inverse variance estimates, with a 1\% floor.  Additionally 25\% and 75\% quantiles of the fluxes in each band are computed.  Given that the typical object has only 3 photometric measurements in each band, this set of summary statistics is unnecessarily extensive, but this pipeline was originally developed for PS1, where more epochs were available in each filter.  An average blendedness (\texttt{fracflux}) is also computed.

Only detections of sources considered ``ok'' are included in the averages.  An ``ok'' detection has no problematic flags (Table~\ref{tab:flags}), and has a ``quality factor'' greater than 0.85 (see \textsection\ref{subsubsec:individualcatalogs}).  A detection with the \texttt{NEBULOSITY} flag set may be considered ``ok,'' since excluding these detections would exclude all measurements of objects in regions with substantial nebulosity.

The object catalog contains entries for slightly over two billion objects.  Table~\ref{tab:mergedcatalog} lists and briefly describes the fields included in the catalog.  

\begin{deluxetable}{ll}
\tablewidth{\columnwidth}
\tablecaption{Merged Catalog Fields}
\tablehead{
\colhead{Name} & \colhead{Description}}
\startdata
\label{tab:mergedcatalog}
\texttt{ra} & right ascension (deg) \\
\texttt{dec} & declination (deg) \\
\texttt{posstdev} & standard deviation in position (\arcsec) \\
\texttt{ra\_ok} & right ascension (ok detections only) (deg) \\
\texttt{dec\_ok} & declination (ok detections only) (deg) \\
\texttt{posstdev\_ok} & standard deviation (ok detections only) (\arcsec) \\
\hline
\texttt{ndet} & number of detections \\
\texttt{ndet\_ok} & number of ok detections \\
\texttt{nmag} & number of detections in each band \\
\texttt{nmag\_ok} & number of okay detections in each band \\
\hline
\texttt{mean} & mean flux (units of 3631 Jy) \\
\texttt{stdev} & flux standard deviation \\
\texttt{err} & uncertainty in mean flux \\
\texttt{median} & median flux (units of 3631 Jy) \\
\texttt{q25} & $25^\mathrm{th}$ percentile flux (units of 3631 Jy) \\
\texttt{q75} & $75^\mathrm{th}$ percentile flux (units of 3631 Jy) \\
\hline
\texttt{epochrange} & MJD between first and last detections \\
\texttt{epochrange\_ok} & MJD between first and last ok detections \\
\texttt{epochmean} & average MJD \\
\texttt{epochmean\_ok} & average MJD (ok only) \\
\hline
\texttt{maglimit} & 5$\sigma$ magnitude limit of deepest image (ok only) \\
\texttt{fracflux} & average fractional flux in each band (ok only) \\
\enddata
\tablecomments{
Fields in the DECaPS merged catalogs.  A more complete description is available at the survey web site.  Most fields are 32-bit floating point numbers, except for fields pertaining to epochs and positions, which are 64-bit floating point numbers.  The fields describing the number of detections of sources are 16-bit integers.
}
\end{deluxetable}

Fluxes in the merged catalogs are given in units of 3631 Jy (equivalently, in Mgy, \citealt{Finkbeiner:2004}).  This is intended to simplify conversion to AB magnitudes; the AB magnitude is $\mathrm{AB} = -2.5\log_{10}\mathrm{flux}$.

Additional statistics about the flux are available in columns labeled \texttt{lbs}, indicating that they refer to local-background-subtracted fluxes (see \textsection\ref{subsubsec:individualcatalogs}).

\section{Conclusion}
\label{sec:conclusion}

The DECam Plane Survey comprises $grizY$ imaging of the southern Galactic plane, $5\degree > l > -120\degree$, $|b| \lesssim 4\degree$, to depths between 24th and 21st mag.  The photometric calibration is accurate to 1--2\%.  Crowded field photometry is used to identify and photometer more than 20 billion detections of 2 billion sources.  Raw and processed images are available online through the NOAO Archive.  Individual image and band-merged object catalogs are available through the survey web site.  DECaPS coadd images, model images, and residuals images are available through the DECam Legacy Survey viewer for rapid inspection.

In combination with PS1, DECaPS provides $grizY$ coverage of 360\degree\ of the Milky Way's midplane.  DECaPS can detect main-sequence turn-off stars out to beyond 8~kpc when the dust reddening $E(B-V) < 1.5$ mag, probing a large portion of the Milky Way.  Existing longer wavelength surveys like VVV, and GLIMPSE are natural counterparts to DECaPS, with less dust-sensitive bands that can see through denser regions of the interstellar medium.

We anticipate that the DECaPS photometry will serve as the foundation for new maps of stars and dust in the southern Galactic plane, complementing maps of the northern Galactic plane like that of \citet{Green:2015}.  The combination of DECaPS photometry and APOGEE-2 spectroscopy \citep{Abolfathi:2017} will similarly enable the shape of the dust extinction curve and its variation to be measured throughout the same volume \citep[e.g.][]{Schlafly:2017}.  Inaccurate dust maps place important limitations on the accuracy that can be achieved by the Gaia mission \citep{Bovy:2016}, motivating deeper photometry and further effort in this area to realize the potential of Gaia's Milky Way studies.  We are excited to learn what other uses the community can find for the billions of stars cataloged by DECaPS.

ES acknowledges support for this work provided by NASA through Hubble Fellowship grant HST-HF2-51367.001-A awarded by the Space Telescope Science Institute, which is operated by the Association of Universities for Research in Astronomy, Inc., for NASA, under contract NAS 5-26555.  GMG and DPF acknowledge support from NSF grant AST-1312891.  Travel to CTIO was partially supported by NSF grant AST-1614941.

DECaPS is based on observations at Cerro Tololo Inter-American Observatory, National Optical Astronomy Observatory (NOAO Prop. ID: 2014A-0429, 2016A-0327, and 2016B-0279; PI: Finkbeiner), which is operated by the Association of Universities for Research in Astronomy (AURA) under a cooperative agreement with the National Science Foundation.

The DECaPS analysis was run on the Odyssey cluster supported by the FAS Division of Science, Research Computing Group at Harvard University, and on the National Energy Research Scientific Computing Center, a DOE Office of Science User Facility supported by the Office of Science of the U.S. Department of Energy under Contract No. DE-AC02-05CH11231.  The DECaPS nebulosity CNN was trained on the XStream computational resource, supported by the National Science Foundation Major Research Instrumentation program (ACI-1429830).

This project used data obtained with the Dark Energy Camera (DECam), which was constructed by the Dark Energy Survey (DES) collaboration.  Funding for the DES Projects has been provided by the U.S. Department of Energy, the U.S. National Science Foundation, the Ministry of Science and Education of Spain, the Science and Technology Facilities Council of the United Kingdom, the Higher Education Funding Council for England, the National Center for Supercomputing Applications at the University of Illinois at Urbana-Champaign, the Kavli Institute of Cosmological Physics at the University of Chicago, the Center for Cosmology and Astro-Particle Physics at the Ohio State University, the Mitchell Institute for Fundamental Physics and Astronomy at Texas A\&M University, Financiadora de Estudos e Projetos, Funda{\c c}{\~a}o Carlos Chagas Filho de Amparo {\`a} Pesquisa do Estado do Rio de Janeiro, Conselho Nacional de Desenvolvimento Cient{\'i}fico e Tecnol{\'o}gico and the Minist{\'e}rio da Ci{\^e}ncia, Tecnologia e Inovac{\~a}o, the Deutsche Forschungsgemeinschaft, and the Collaborating Institutions in the Dark Energy Survey.  The Collaborating Institutions are Argonne National Laboratory, the University of California at Santa Cruz, the University of Cambridge, Centro de Investigaciones En{\'e}rgeticas, Medioambientales y Tecnol{\'o}gicas-Madrid, the University of Chicago, University College London, the DES-Brazil Consortium, the University of Edinburgh, the Eidgen{\"o}ssische Technische Hoch\-schule (ETH) Z{\"u}rich, Fermi National Accelerator Laboratory, the University of Illinois at Urbana-Champaign, the Institut de Ci{\`e}ncies de l'Espai (IEEC/CSIC), the Institut de F{\'i}sica d'Altes Energies, Lawrence Berkeley National Laboratory, the Ludwig-Maximilians Universit{\"a}t M{\"u}nchen and the associated Excellence Cluster Universe, the University of Michigan, {the} National Optical Astronomy Observatory, the University of Nottingham, the Ohio State University, the University of Pennsylvania, the University of Portsmouth, SLAC National Accelerator Laboratory, Stanford University, the University of Sussex, and Texas A\&M University.

The Pan-STARRS1 Surveys (PS1) and the PS1 public science archive have been made possible through contributions by the Institute for Astronomy, the University of Hawaii, the Pan-STARRS Project Office, the Max-Planck Society and its participating institutes, the Max Planck Institute for Astronomy, Heidelberg and the Max Planck Institute for Extraterrestrial Physics, Garching, The Johns Hopkins University, Durham University, the University of Edinburgh, the Queen's University Belfast, the Harvard-Smithsonian Center for Astrophysics, the Las Cumbres Observatory Global Telescope Network Incorporated, the National Central University of Taiwan, the Space Telescope Science Institute, the National Aeronautics and Space Administration under Grant No. NNX08AR22G issued through the Planetary Science Division of the NASA Science Mission Directorate, the National Science Foundation Grant No. AST-1238877, the University of Maryland, Eotvos Lorand University (ELTE), the Los Alamos National Laboratory, and the Gordon and Betty Moore Foundation.

\facility{Blanco (DECam)}

\appendix

\section{Nebulosity Network Structure}
\label{app:nebulosity-network-structure}

Our convolutional neural network takes histogram-normalized $512 \times 512$-pixel images as input, selected from the CP \texttt{InstCal} stage images. The network consists of 14 convolutional layers, interspersed with 6 maximum pooling layers. A global average pooling layer reduces the activations of the last maximum pooling layer to 32 activations, each representing a different learned feature in the input image. These 32 features are finally fed into a two-layer dense neural network, which classifies each image as one of our four hand-classified image types: \texttt{NEBULOSITY}, \texttt{NEBULOSITY\_LIGHT}, \texttt{NORMAL}, or \texttt{SKY\_ERROR}.

Each convolutional and dense layer is followed by a ReLU activation layer \citep{Hahnloser:2000}. We use categorical cross-entropy as our loss function. In all, our neural network has 75352 trainable parameters. In order to avoid over-fitting, we use L2 weight regularization in the convolutional layers and dropout in the dense layers.

Table~\ref{tab:network-architecture} summarizes our network architecture.

\begin{deluxetable}{c|c|c}
    \tablecaption{Network architecture}
    \tablehead{\colhead{layer} & \colhead{output shape} & \colhead{details}
    \label{tab:network-architecture}
    }
    \startdata
        conv2d\_1 &
        $512 \times 512 , \ 12$ &
        $5 \times 5 , \, \mathrm{same \ padded}$ \\
        maxpool2d\_1 &
        $256 \times 256 , \ 12$ &
        $2 \times 2$ \\ \hline
        conv2d\_2 &
        $256 \times 256 , \ 24$ &
        $5 \times 5 , \, \mathrm{same \ padded}$ \\
        maxpool2d\_2 &
        $128 \times 128 , \ 24$ &
        $2 \times 2$ \\ \hline
        conv2d\_3 &
        $128 \times 128 , \ 24$ &
        $\begin{pmatrix}
        3 \times 3 , \, \mathrm{same \ padded} \\
        3 \times 3 , \, \mathrm{same \ padded} \\
        1 \times 1 \hphantom{, \, \mathrm{same \ padded}}
        \end{pmatrix}$ \\
        maxpool2d\_3 &
        $64 \times 64 , \ 24$ &
        $2 \times 2$ \\ \hline
        conv2d\_4 &
        $64 \times 64 , \ 32$ &
        $\begin{pmatrix}
        3 \times 3 , \, \mathrm{same \ padded} \\
        3 \times 3 , \, \mathrm{same \ padded} \\
        1 \times 1 \hphantom{, \, \mathrm{same \ padded}}
        \end{pmatrix}$ \\
        maxpool2d\_4 &
        $32 \times 32 , \, 32$ &
        $2 \times 2$ \\ \hline
        conv2d\_5 &
        $32 \times 32, \ 32$ &
        $\begin{pmatrix}
        3 \times 3 , \, \mathrm{same \ padded} \\
        3 \times 3 , \, \mathrm{same \ padded} \\
        1 \times 1 \hphantom{, \, \mathrm{same \ padded}}
        \end{pmatrix}$ \\
        maxpool2d\_5 &
        $16 \times 16 , \ 32$ &
        $2 \times 2$ \\ \hline
        conv2d\_6 &
        $12 \times 12, \ 32$ &
        $\begin{pmatrix}
        3 \times 3 \\
        3 \times 3 \\
        1 \times 1
        \end{pmatrix}$ \\
        maxpool2d\_6 &
        $6 \times 6 , \ 32$ &
        $2 \times 2$ \\ \hline
        global\_avg\_pool2d &
        32 &
        \\ \hline
        dense\_1 &
        12 &
        20\% dropout \\
        dense\_2 &
        4 &
        10\% dropout \\ \hline
        softmax &
        4 & \\
        \enddata
    \tablecomments{All convolutional and dense layers use ReLU activation. The final output one-hot encodes the class, and the cross-entropy loss function is used.}
\end{deluxetable}

\bibliography{2dmap}

\end{document}